\documentclass{article}
\usepackage{epsfig}
\usepackage{color}
\usepackage{amsmath}
\usepackage{amssymb}
\usepackage{lscape}
\usepackage{ulem}

\makeatletter
\@addtoreset{equation}{section}

\makeatother

\begin{document}

\fontsize{11pt}{0.5cm}\selectfont

\title{
\vbox{
\baselineskip 14pt
\hfill \hbox{\normalsize KEK-TH-2508
}} \vskip 1cm
\bf \Large  
Stringy Threshold Corrections in D-brane Systems 
\vskip .5cm
}

\author{
Satoshi Iso$^{a,b,c}$  \thanks{E-mail: \tt iso(at)post.kek.jp}, 
Noriaki Kitazawa$^{d}$ \thanks{E-mail: \tt noriaki.kitazawa(at)tmu.ac.jp},
Takao Suyama$^{a}$ \thanks{E-mail: \tt tsuyama(at)post.kek.jp}
\bigskip\\
\it \normalsize
$^a$ Theory Center, Institute for Particles and Nuclear Studies (IPNS), \\
\it  \normalsize 
High Energy Accelerator Research Organization (KEK), \\
\it  \normalsize 
\it \normalsize Oho 1-1, Tsukuba, Ibaraki 305-0801, Japan \\
\it  \normalsize 
$^b$ 
International Center for Quantum-field Measurement Systems \\
\it \normalsize for Studies of the Universe and Particles (QUP),  KEK \\
\it \normalsize  Oho 1-1, Tsukuba, Ibaraki 305-0801, Japan  \\
\it  \normalsize 
$^c$Graduate University for Advanced Studies (SOKENDAI),\\
\it \normalsize Oho 1-1, Tsukuba, Ibaraki 305-0801, Japan \\
\it  \normalsize 
$^d$Department of Physics, Tokyo Metropolitan University,\\
\it  \normalsize Hachioji, Tokyo 192-0397, Japan \\
\smallskip
}
\date{\today}

\maketitle

\begin{center}
\bf Abstract
\end{center}

We investigate string amplitudes by using the {\it partial modular transformation} which we introduced in our previous works. 
This enables us to extract stringy threshold corrections from the full string amplitudes and interpret them  in terms of the Wilsonian  effective field theory  in a natural way. 
We calculate mass shifts and wave function renormalizations for massless scalar fields on brane-antibrane systems. 
We find that the mass shift can be exponentially small and negative. 
We also propose a strategy for realizing a hierarchical mass spectrum on D-branes. 

\newpage

\vspace{1cm}

\section{Introduction}

\vspace{5mm}

String theory is one of promising candidates of  unification of matter and spacetime
and expected to solve various problems in the standard model of particle physics. 
A conventional approach towards the goal is to construct a string vacuum and analyze the dynamics of 
 its low-energy (massless) spectrum. Such a treatment will be sufficient for many purposes but sometimes 
 may miss some important characteristics of string theory. 
 For example, string theory contains an infinite tower of 
 massive states and their  collective behavior is known to 
 drastically soften the ultraviolet (UV) behaviors, such as  the high energy scattering amplitude, 
and makes the string theory UV finite. 
It is reasonable to expect that similar miracles would happen also in low-energy observables. 

In ordinary quantum field theories, we have two types of UV divergences which can be controlled by 
the renormalization procedure; power divergences and logarithmic ones.  
The latter divergences are reflected by the running
(scale dependence) of physical quantities  and the behavior is observable  at the low-energy scale. 
The former types of divergences include 
quartic divergence of the cosmological 
constant and quadratic divergence of the Higgs mass term in (3+1) dimensions.
Once they are absorbed into the bare parameters at UV cutoff,
they never appear in the low-energy effective theory; they are not observables.  
But  in the process of absorbing, fine tuning is necessary to obtain sensible low-energy observables 
such as  125 GeV Higgs mass or meV dark energy in our universe
against  UV cutoff scales. 
This problem of fine tuning is called the naturalness problem in the standard model of particle physics. 
There are various approaches to the naturalness problem within the framework of ordinary quantum field
theories, but none has successfully given a solution. 
  
The difficulty in solving this problem seems to suggest that we will need to go beyond the framework of quantum field theory
and treat both of matter and gravity on an equal footing. 
An example of such an approach is the misaligned supersymmetry
  \cite{Dienes:1994np,Dienes:1995pm}. 
In these papers, the authors have shown that power divergences can be weakened due to 
the cooperation of infinite tower of states.
Though no models are found to remove all the power divergences, 
 it is interesting that we can soften the UV behavior by  intrinsic properties of string theory such as modular invariance. 
In \cite{Abel:2021tyt,Abel:2023hkk}, the string threshold corrections to the Higgs mass 
and running gauge coupling are calculated by summing all the infinite towers of massive states. 

Our discussion in this paper is along the same line of thought as above. 
We consider D-brane systems in string theory as a UV complete framework of  models of particle physics.
Our analysis is motivated by our expectation that  stringy effects may  realize a hierarchical structure of scales between the low-energy scales and the string scale. 
We also utilize the modular transformation property 
in calculating  mass corrections  in  string theory. 
We particularly use  the {\it partial modular transformation}, which
enables us to separate the radiative corrections in field theory and the threshold corrections from string theory
in a systematic way. 

\vspace{5mm}

In this paper, we calculate  quantum shifts of masses for massless fields in the low-energy effective theory of string theory. 
In quantum field theories, the calculation of the mass shift is a standard procedure. 
Namely, we extract the mass shift from an off-shell 2-point amplitude. 
This procedure cannot be straightforwardly applied to string theory since the perturbative formulation of string theory tells us only constructions of S-matrices, not of Green functions. 

Moreover, since the tree level masses are determined completely by the conformal invariance, any shift of the masses will spoil the conformal invariance, and the resulting amplitudes would depend on the worldsheet coordinates chosen to perform the actual calculation. 
This issue was clarified in \cite{Weinberg:1985je,Seiberg:1986ea} as follows. 

\begin{figure}
\begin{center}
\includegraphics{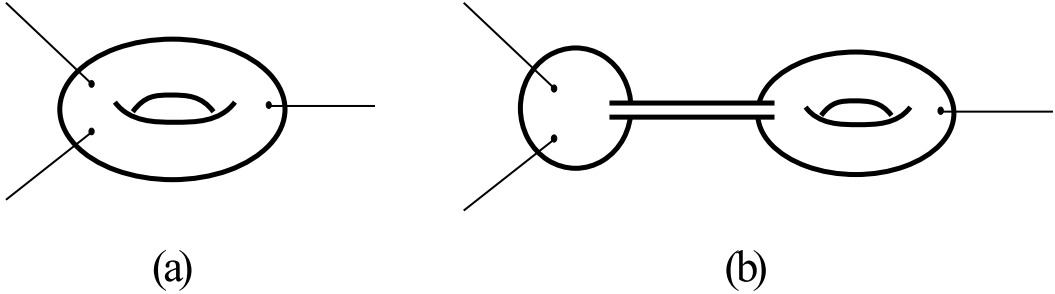}
\end{center}
\caption{(a) A torus with three vertex operators inserted. 
(b) The worldsheet conformally equivalent to the limit of (a) when left two vertex operators come close to each other. }
   \label{fig_Riemann_surface}
\end{figure}

Consider a 3-point amplitude at 1-loop level in closed string theory. 
The worldsheet for this amplitude is depicted in figure \ref{fig_Riemann_surface}(a). 
The positions of the vertex operators are integrated over the worldsheet. 
For a corner of the integration region in which two of the three vertex operators come close to each other, the worldsheet is conformally equivalent to the one depicted in figure \ref{fig_Riemann_surface}(b). 
The long tube in the middle corresponds to  propagations of closed string states. 
By the momentum conservation, the momentum flowing in the tube is on-shell, resulting in a divergence of the amplitude. 
In quantum field theories, this divergence is remedied by the mass shift. 
In \cite{Seiberg:1986ea}, this divergence is interpreted as anomalous dimensions of the vertex operators. 
Then, in order to preserve the conformal invariance even when 1-loop contributions are included, we should modify the mass-shell conditions of the external momenta so that the vertex operators including the anomalous dimensions have conformal dimension 1. 
As a result, the tree-level masses receive 1-loop corrections which are given by certain 2-point 1-loop amplitudes. 
This is similar to the Fischler-Susskind mechanism
 \cite{Fischler:1986ci,Fischler:1986tb}. 

The argument in \cite{Weinberg:1985je,Seiberg:1986ea} can be interpreted in terms of the BRST anomaly \cite{Sen:1987bd}. 
Suppose that we consider an amplitude including the vertex operator of a null state. 
The null state is BRST exact, and therefore the amplitude must vanish. 
However, there are possibilities that non-zero contributions might come from boundaries of the moduli space of the worldsheet. 
The points on the boundary of the moduli space correspond to degenerate worldsheets like the one in figure \ref{fig_Riemann_surface}(b). 
The contribution to the amplitude from the null state can then be cancelled by a prescribed change in the vertex operators due to the shift of the masses. 
This argument can be extended to higher genus amplitudes \cite{Sen:1987bd}. 
See also \cite{Das:1988fu,Rey:1988us}. 

Based on the investigations reviewed above, it has been established that the mass shift including threshold corrections due to all massive states can be obtained from a suitable 2-point string amplitude. 
Following this prescription, mass shifts have been calculated in, for example, 
\cite{Yamamoto:1987dk,Tsuchiya:1988ps,Amano:1988ht,Minahan:1989cb,Kitazawa:2006if,DelDebbio:2008hb,Chialva:2009pg,Kashyap:2023dnk,Eberhardt:2023xck}. 
See also \cite{Sundborg:1988ai,Yamaguchi:1989am,Sundborg:1989jv,Antoniadis:2000tq,Antoniadis:2001cv,Kitazawa:2008tb,Kitazawa:2012hr}. 

\vspace{5mm}

The relation between mass shifts and 2-point amplitudes explained above matches with our intuition developed from quantum field theory, but the argument relating them looks rather indirect. 
The main reason for this is, as mentioned above, the difficulty in off-shell formulation of string theory. 
The issue should be understood more straightforwardly based on string field theory. 
However, string field theory is not always available, especially for superstring theories. 

Remarkably, there has been some progress in understanding off-shell amplitudes in string theory without relying on string field theory. 
As mentioned above, the difficulty in extending perturbative string amplitudes to off-shell states is the following. 
Suppose that we simply make the external momenta in the vertex operators off-shell. 
We call the resulting quantity a naive extension. 
Since the vertex operators do not have the correct conformal dimensions, the conformal invariance is broken, and the naive extension depends on the choice of local coordinates on the worldsheet. 
This makes the definition of the off-shell amplitudes ambiguous. 

In \cite{Pius:2013sca,Pius:2014iaa}, it is shown that such naive extensions, although coordinate dependent, still have important physical information, as long as the local coordinates are chosen appropriately. 
The condition required for the local coordinates is called ``gluing compatibility'' in \cite{Pius:2014iaa} which roughly means that the local coordinates should be available even in the singular limits of the worldsheet, like in figure \ref{fig_Riemann_surface}(b). 
Assuming the gluing compatibility, it is shown that the quantum-corrected masses read off from the positions of poles of the naive extension turn out to be independent of the coordinate choice. 
For the S-matrix, the coordinate dependence of the naive extension turns out to be cancelled by the coordinate dependence of the wave-function renormalization \cite{Pius:2013sca,Pius:2014iaa}. 
In fact, the validity of the naive extension was already observed in some sense for 1-loop amplitudes \cite{Bern:1992ad,DHoker:1993hvl,DHoker:1993vpp,DHoker:1994gnm,DiVecchia:1996uq}. 
In light of \cite{Pius:2013sca,Pius:2014iaa}, this is because we can choose global coordinates, up to a periodic identification, on the torus and the annulus, and it trivially satisfies the gluing compatibility condition. 

There is also a recent attempt for an off-shell formulation of string theory \cite{Ahmadain:2022tew,Ahmadain:2022eso}. 

\vspace{5mm}

Once the off-shell extension of string amplitudes becomes available, we can understand the quantum corrections in string theory much more intuitively. 
In this paper, we provide an explicit argument for relating 2-point amplitudes to the mass shift and the wave function renormalization. 
The essence of the argument is as follows. 

We consider 2-point amplitudes of open string states living on some D-brane systems. 
At 1-loop level, the 2-point amplitudes have expressions schematically of the form 
\begin{equation}
{\cal A}\ =\ \int_0^\infty dt\,I(t), 
\end{equation}
where $t$ is the modulus of the annulus. 
We divide this into two parts: 
\begin{equation}
{\cal A}\ =\ \int_{\Lambda^{-2}}^\infty dt\,I(t)+\int_{\Lambda^2}^\infty ds\,\tilde{I}(s), 
   \label{PMT_intro}
\end{equation}
where $s:=1/t$ and $\tilde{I}(s)$ is the modular transform of $I(t)$ multiplied by $s^{-2}$. 
This procedure is called the partial modular transformation in \cite{Iso:2019gbd,Iso:2020ewj} where we used it to investigate an interaction bewteen moving D-branes. 
Note that the same idea also appeared in \cite{Douglas:1996yp} in a different context. 

The utility of the partial modular transformation is as follows. 
We will show in examples that the first term in (\ref{PMT_intro}) gives the 2-point {\it off-shell} amplitude of the low-energy effective theory, where $t$ is the Schwinger parameter and $\Lambda$ is a UV cut-off. 
This confirms that the naive off-shell extension of string amplitudes is valid at least at 1-loop level. 
The second term, originally corresponding to small Schwinger parameter $t$, is then interpreted as corrections to the effective theory amplitude coming from integrating out high energy modes. 
Indeed, we will show that the second term, which originally includes an integral of the position of vertex operators, can be written as if they are contributions from local higher-dimensional operators. 
In other words, the expression (\ref{PMT_intro}) can be interpreted as the amplitude obtained from the Wilsonian effective action of string theory with cut-off $\Lambda$. 
Based on this interpretation, it is now apparent that the mass shift as well as the wave function renormalization can be obtained from the 2-point string amplitude. 
Therefore, the evaluation of the 2-point string amplitudes amounts to taking into account threshold corrections due to all massive states circulating in the loop. 

\vspace{5mm}

\begin{figure}
\begin{center}
\includegraphics{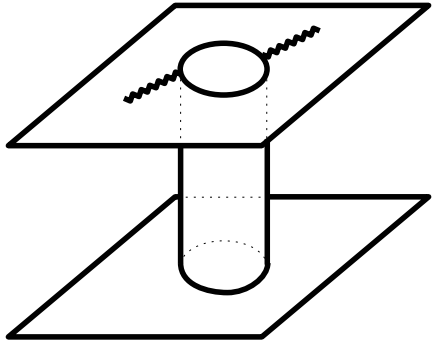}
\end{center}
\caption{Typical setup we consider in this paper. 
There are two D-branes. 
We consider 1-loop string amplitudes of open string states, represented by wavy lines, living on the upper D-brane. 
An open string stretched between the D-branes, represented by cylinder, is circulating in the loop.}
   \label{fig_1-loop_planar}
\end{figure}

So far in the literature, the mass shift has been calculated mainly for massive states in superstring theories. 
This is simply because the massless states are protected by symmetries. 
In this paper, we consider quantum corrections to open string states on D-branes which are massless at tree level, but not necessarily protected by symmetries. 
We choose the setup, depicted in figure \ref{fig_1-loop_planar}, in which the supersymmetry is completely broken. 
Therefore, the scalars corresponding to the transverse directions of the D-branes may acquire non-zero masses at 1-loop. 
Note that originally these scalars on a D-brane are massless due to the fact that they are Nambu-Goldstone bosons for the translational symmetry. 
In presence of other D-branes, the translational symmetry is broken, allowing the scalars to become massive. 

In general, an amplitude in this setup may suffer from the problem of divergences. 
For example, the open string stretched between the two D-branes gives a tachyon which causes an infrared (IR) divergence. 
We avoid this by keeping distance of the D-branes sufficiently long so that the negative zero-point energy is cancelled by the string tension. 
Another source of divergence is the dilaton tadpole. 
Indeed, if the D-branes are space-filling, then the presence of a non-zero dilaton tadpole causes a divergence. 
This is remedied simply by focussing our attension on lower-dimensional D-branes\footnote{
The divergence for space-filling D-branes is due to the fact that the propagator $1/q^2$ of the dilaton is evaluated at $q=0$ due to the momentum conservation. 
For lower dimensional D-branes, the transverse momentum does not need to be conserved, and the momentum integration gives us a finite value corresponding to the effective potential, as long as the D-branes are separated. 
}. 
The presence of a closed string tachyon is a fatal catastroph. 
We avoid this by keeping the supersymmetry in the bulk, or otherwise we simply subtract the closed string tachyon by hand. 
Taking them into account, we will show completely well-defined calculations of 1-loop amplitudes in section \ref{sec:typeII}. 

The amplitudes for the above setup provide us with interesting results, due to the presence of one parameter, namely the distance $r$ between the D-branes. 
As a result, the quantum corrections are functions of $r$. 
We show that, for example, the mass shift of the transverse scalar is decomposed into a sum of the following form 
\begin{equation}
\Delta m^2(r)\ =\ \Delta m^2_{(0)}(r)+\sum_{n=1}^\infty \Delta m^2_{(n)}(r), 
\end{equation}
where $\Delta m^2_{(0)}(r)$ is proportional to a negative power of $r$, while $\Delta m^2_{(n)}(r)$ with $n\ge1$ are exponentially discreasing with $r$. 
This result can be explained easily from the viewpoint of the partial modular transformation. 
Namely, it is because the quantum corrections are given by closed string exchanges between the D-branes. 
The situation is similar to the calculation of an interaction between two D-branes via the cylinder amplitude.
A difference is that in our case two vertex operators are inserted to one of the boundaries of the cylinder. 
The exchanges of massless closed string states give $\Delta m^2_{(0)}(r)$, just like similar exchange amplitudes 
give the Newtonian potential in case of the effective potential between D-branes. 
On the other hand, the exchanges of massive closed string states give $\Delta m^2_{(n)}(r)$ with $n\ge1$ corresponding to the Yukawa potential. 
The wave function renormalization also has the same structure. 

This result suggests an interesting possibility to realize a large hierarchy in the mass spectrum. 
Suppose that one finds a D-brane setup for which $\Delta m^2_{(0)}(r)$ vanishes. 
This is possible when the contributions from massless closed string exchanges cancel among them. 
If it is the case, then the leading order correction to the mass is given by the second term
$\Delta m^2_{(1)}(r)$ which is vanishingly small even when $r$ is, say, only 5 times the string scale. 
In this way, the massless scalars could acquire a non-zero mass far smaller than the string scale. 
It is very interesting to notice that this mechanism crucially relies on the open-closed duality, a genuine stringy effect. 

\vspace{5mm}

This paper is organized as follows. 
In section \ref{sec:PMT}, we review the partial modular transformation, and then apply it to a 2-point 1-loop string amplitude in bosonic string theory. 
We will see that the string amplitude can be approximated by the corresponding amplitudes in gauge theory and gravity with additional certain stringy corrections. 
In section \ref{sec:renorm}, we argue a relation between the partial modular transformation and the Wilsonian renormalization by introducing a floating division point $\lambda$ in the partial modular transformation. 
We discuss the structure of the mass shift and the wave function renormalization in a simple setup of bosonic string theory. 
The results have ambiguities due to the closed string tachyon. 
This problem is remedied in section \ref{sec:typeII} where we discuss D-brane systems in Type II string theory. 
In particular, we will show that the mass shifts of the transverse scalars of the D-brane can be exponentially small and negative in a particular case. 
Section \ref{sec:discuss} is devoted to discussions. 

There are various appendices for technical details. 
In appendix \ref{app_amplitude}, we briefly review 2-point 1-loop string amplitudes discussed in this paper. 
Appendix \ref{app_modular} contains formulas for the modular transformation of various functions. 
In this paper, we need to compare the string amplitudes with the corresponding field theory amplitudes represented by using the Schwinger parameters. 
The latter amplitudes are summarized in appendix \ref{app_QFT}. 
Appendix \ref{app_leading} contains detailed calculations of the string amplitudes expanded in terms of the external momentum. 
Some functions which are used in analyzing the structure of the mass shift are introduced in appendix \ref{app_bulk}. 
We employ the Green-Schwarz formalism in calculating the string amplitudes in Type II string theory. 
For this calculation, we need to specify the boundary condition of open strings attached to D-branes. 
We determine the boundary condition in appendix \ref{app_GS}. 
We show in appendix \ref{app_DBI} a calculation corresponding to the wave function renormalization, based on Type II supergravity and DBI action. 
Another different type of D-brane setup which is a modification of the one investigated in section \ref{sec:typeII} is discussed in appendix \ref{modified_SS}. 

\vspace{1cm}

\section{String amplitudes and partial modular transformation} \label{sec:PMT}

\vspace{5mm}

In this section, we apply the partial modular transformation introduced in \cite{Iso:2019gbd,Iso:2020ewj} to string amplitudes. 
We will show that the partial modular transformation allows us to expand the amplitudes in terms of a small parameter. 
The leading order terms in this expansion are given in terms of gauge theory and gravity, 
with extra terms which are interpreted as ``stringy corrections.'' 
In order to perform the calculations, we need a regularization since the moduli integration of the string amplitude is divergent when the vertex operators come close to each other. 
We will show its validity
by checking that the regularized results are consistent with the gauge symmetry of the worldvolume theory of D-branes. 

\vspace{5mm}

\subsection{Effective potential} \label{subsec_effective_potential}

\vspace{5mm}

First, we recall the partial modular transformation applied to an effective potential for a pair of D-branes  \cite{Iso:2019gbd,Iso:2020ewj}. 
As an illustrative example, consider a D$p$-brane and a $\overline{{\rm D}p}$-brane in Type II string theory. 
We denote the distance between the D-branes by $l$. 
The effective potential at the leading order in the string coupling is given by the cylinder amplitude. 
The explicit expression is 
\begin{equation}
V(r)\ =\ \int_0^\infty\frac{dt}t\,I(t), \hspace{1cm} I(t)\ :=\ (8\pi^2\alpha^\prime  t)^{-(p+1)/2}e^{-2\pi r^2t}\frac{\vartheta_{01}(0,it)^4}{\eta(it)^{12}}, 
   \label{effective_potential}
\end{equation}
where $t$ is the circumference of the cylinder and 
\begin{equation}
r^2\ :=\ \frac{l^2}{4\pi^2\alpha^\prime}.
\end{equation}
Our notations of the eta function and theta functions follow those in section 7.2 in \cite{Polchinski:1998rq}. 
For convenience of the readers, their definitions are shown in appendix \ref{app_modular}. 
The part $(8\pi^2\alpha^\prime  t)^{-(p+1)/2}$ comes from the momentum integration along the D-brane, 
$e^{-2\pi r^2t}$ represents the effect of the string tension due to the separation of D-branes, and the other parts in $I(t)$ come from the non-zero modes of the open string between D-branes. 
The effective potential $V(r)$ contains contributions from all massive open string modes circulating along the cylinder. 
Because of this, $V(r)$ has a rather complicated expression. 

The properties of the effective potential can be understood more easily by using the partial modular transformation. 
It consists of two steps: the first step is to divide the $t$-integral into IR region $(t > 1)$ and UV region $(t \le 1)$. 
The second step is to perform the modular transformation only for the UV region. 
The result for $V(r)$ is 
\begin{equation}
V(r)\ =\ \int_1^\infty \frac{dt}t\,I(t)+\int_1^\infty ds\,\tilde{I}(s), 
   \label{PMT_potential}
\end{equation}
where $s=1/t$, and 
\begin{equation}
\tilde{I}(s)\ =\ (8\pi^2\alpha^\prime  )^{-(p+1)/2}s^{(p-9)/2}e^{-2\pi r^2/s}\frac{\vartheta_{10}(0,is)^4}{\eta(is)^{12}}. 
\end{equation}
Here we have used the modular transformation properties
\begin{equation}
\eta(it)\ =\ s^{1/2} \eta(is), \hspace{1cm} \vartheta_{01}(0, it) = s^{1/2} \vartheta_{10}(0,is).
\end{equation}
The utility of this transformation is the following. 
The integrands $I(t)$ and $\tilde{I}(s)$ depend on $q:=e^{-2\pi t}$ and $\tilde{q}:=e^{-2\pi s}$, respectively. 
Since the integration regions in (\ref{PMT_potential}) have a lower bound $t,s\ge1$, these quantities are bounded from above as 
\begin{equation}
q,\tilde{q}\ \le\ e^{-2\pi}\ \sim\ 0.00187. 
\end{equation}
Therefore, the $q$-expansion for $I(t)$ and the $\tilde{q}$-expansion for $\tilde{I}(s)$ effectively gives an expansion of $V(r)$ in terms of a small parameter $e^{-2\pi}$. 

According to this expansion, we can approximate the first integral in $V(r)$ as 
\begin{equation}
\int_1^\infty \frac{dt}t\,I(t)\ \sim\ \int_1^\infty \frac{dt}t\,(8\pi^2\alpha^\prime  t)^{-(p+1)/2}e^{-2\pi(r^2-1/2)t}
\end{equation}
up to terms of order ${\cal O}(q^{1/2})$. 
As  stringy states corresponding to higher powers of $q$ are dropped, 
the right-hand side coincides with the vacuum 1-loop amplitude in the Schwinger parametrization of a scalar field in $p+1$ dimensions with mass $m^2=(r^2-\frac12)/\alpha^\prime$. 
Note that a UV cut-off of the string scale corresponding to $t=1$ is introduced. 
This scalar field comes from the ground state of the open string stretched between the
  D$p$ and $\overline{{\rm D}p}$-branes. 
This is not tachyonic as long as the distance $r$ is large enough. 

The second integral in $V(r)$ can be approximated as 
\begin{equation}
\int_1^\infty ds\,\tilde{I}(s)\ \sim\ 16(8\pi^2\alpha^\prime  )^{-(p+1)/2}\int_1^\infty ds\,s^{(p-9)/2}e^{-2\pi r^2/s}. 
\end{equation}
The right-hand side is the amplitude for the exchange of the massless closed string states between the D-branes, where a short distance cut-off is introduced. 
Since we consider a brane-antibrane system, the contributions from NS-NS states and R-R states add up. 
If the cut-off is removed, then the right-hand side is proportional to $r^{p-7}$, giving the Newtonian potential between the D-branes. 

The modular transformation is conventionally used to reveal the long distance behavior of the potential between
D-branes from the open string one-loop amplitude. 
If we use the partial modular transformation and sum up two contributions of $I(t)$ and $\tilde{I}(s)$, 
we can combine the short  and 
long distance behaviors to obtain an effective potential which interpolates between small $r$ and large $r$ regions. 

This example shows that the effective potential can be nicely approximated for all the region of  $r>1/\sqrt{2}$ by a sum of the corresponding amplitudes in the D-brane worldvolume theory and in supergravity with the DBI action. 
If one naively added two amplitudes from field theory and gravity, then the resulting expression would not make sense. 
This is because, for example, the UV contributions in gauge theory is doubly counted as the long distance contributions in gravity. 
To avoid such a double-counting, we need to introduce suitable cut-offs both for the field theory amplitudes and the gravity amplitudes, as is given by the partial modular transformation. 

\vspace{5mm}

This observation indicates that the partial modular transformation can be a nice approximation scheme for calculating the effective potential for various D-brane systems. 
In particular, this can be applied to complicated D-brane systems for which the quantization of the relevant open string is difficult, provided that the necessary calculations in field theory and gravity can be performed. 
Indeed, in \cite{Iso:2019gbd,Iso:2020ewj} we applied the technique of the 
partial modular transformation to a system of revolving D-branes, and obtained the leading contributions to the effective potential 
for all the region of $r$ in the sense explained above. 

Note that the division of the integral at $t=1$ is an optimal choice for the expansion 
of $V(r)$. 
If the division point is much smaller or much larger, then either the $q$-expansion of $I(t)$ or the $\tilde{q}$-expansion of $\tilde{I}(s)$ gives a worse approximation of their contributions to $V(r)$. 
In section \ref{sec:renorm}, we will show that the change of the division point will have an interesting physical meaning as the Wilsonian renormalization scale. 

\vspace{5mm}

\subsection{String amplitudes}
\label{sec:PMTstringamplitude}
\vspace{5mm}

It would be natural to expect that a similar approximation can be performed for string amplitudes by applying the 
technique of the partial modular transformation. 
This is a very interesting possibility because it may provide a convenient method to calculate the stringy 
threshold corrections to various quantities in the low-energy effective theory of string theory. 
Recall that the method of the partial modular transformation
enables us to calculate an approximate form of the effective potential, 
including contributions from all massive open string modes for all the region of $r$,
in terms of a sum of contributions from the
 field theory on the world volume and gravity in the bulk. 
Then we expect that 
string amplitudes may also have similar  approximations. 
It turns out, however, that the situation is more complicated.
We will show that string amplitudes are not well approximated by only a sum of 
the gauge theory and gravity contributions, but additional  ``stringy threshold corrections'' are necessary to be
included. 
One of the purposes of this section is to calculate such corrections. 

\vspace{5mm}

In the rest of this section, 
in order to grasp the behavior of string amplitudes mentioned above, 
we first consider a simple system of
two D$p$-branes in bosonic string theory, 
and look at the IR ($t \ge 1$) and UV ($t \le 1$) parts of 2-point string amplitude
with gauge bosons as  the external states.
The setup is the same as the one depicted in figure \ref{fig_1-loop_planar}.  
Let $\epsilon_1^\mu,\epsilon_2^\mu$ be their polarization vectors, and $k_1^\mu=k^\mu$ and $k_2^\mu=-k^\mu$ be their external momenta. 
They satisfy the transversality conditions $k_1\cdot\epsilon_1=k_2\cdot\epsilon_2=0$. 
Based on the reason explained in the introduction, we will take $k^\mu$ to be off-shell. 

The calculation of the 1-loop amplitude of figure \ref{fig_1-loop_planar} is given in (\ref{app_amplitude_explicit}) 
as 
\begin{equation}
{\cal A}\ =\ C\epsilon_1\cdot\epsilon_2\int_0^\infty dt\int_0^t d\nu\,I_0(\nu,t)I_1(\nu,t), 
\end{equation}
where $\nu$ is the distance between the vertex operators on the worldsheet which is integrated over the region $0\le \nu\le t$. 
The integrand is a product of the two quantities given by 
\begin{eqnarray}
I_0(\nu,t) 
&:=& (8\pi^2\alpha^\prime  t)^{-(p+1)/2}e^{-2\pi r^2t}\eta(it)^{-24}\exp\left[ -2\alpha^\prime  k^2\left( \pi\frac{\nu(t-\nu)}t+G_B(\nu,t) \right) \right], 
   \label{integrand_gauge_IR-1} \nonumber \\ \\ [2mm] 
I_1(\nu,t) 
&:=& \frac{2\pi}t-\partial_\nu^2G_B(\nu,t), 
   \label{integrand_gauge_IR-2}
\end{eqnarray}
where 
\begin{equation}
G_B(\nu,t)\ :=\ \log\left[ (1-e^{-2\pi\nu})\prod_{m=1}^\infty \frac{(1-e^{-2\pi\nu}q^m)(1-e^{2\pi\nu}q^m)}{(1-q^m)^2} \right], \hspace{1cm} q = \ e^{- 2 \pi t}
   \label{def_G_B}
\end{equation}
is the non-zero mode part of the scalar propagator on the worldsheet. 
The insertion of two vertex operators are reflected in the exponential part of  
$I_0(\nu,t)$ in (\ref{integrand_gauge_IR-1}) and $I_1(\nu,t)$ in (\ref{integrand_gauge_IR-2}). 
For details of the derivation, see appendix \ref{app_amplitude}. 
The overall coefficient $C$ will be fixed later. 

\begin{figure}
\begin{center}
\includegraphics{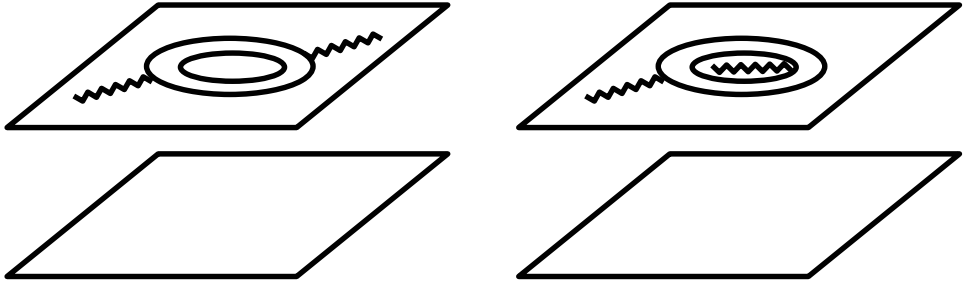}
\end{center}
\caption{String diagrams which do not contribute to the 2-point 1-loop amplitude for the gauge bosons on the upper D$p$-branes since they cancel each other. }
   \label{fig_1-loop_irrelevant}
\end{figure}

Note that there are other string diagrams depicted in figure \ref{fig_1-loop_irrelevant} which are of the same order with respect to the string coupling constant to the above amplitude. 
Actually, those diagrams cancel each other since the open string, whose both ends are attached to the same D-brane, is neutral for the gauge field under consideration. 
This is fortunate for us since the open string tachyon circulating in these loops cannot be eliminated in a simple manner. 

\vspace{5mm}

Let us apply the partial modular transformation to ${\cal A}$. 
The amplitude is divided as 
\begin{equation}
{\cal A}\ =\ {\cal A}_{\rm IR}+{\cal A}_{\rm UV}. 
\end{equation}
The IR part of ${\cal A}$ is defined by restricting the modulus parameter integration into  $t \in [1, \infty)$ and given by
\begin{equation}
{\cal A}_{\rm IR}\ :=\ C\epsilon_1\cdot\epsilon_2\int_1^\infty dt\int_0^t d\nu\,I_0(\nu,t)I_1(\nu,t).
\end{equation}
The other part, corresponding to the UV region $t \in [0, 1]$ of the world volume theory, can be rewritten by the modular transformation as 
\begin{equation}
{\cal A}_{\rm UV}\ :=\ C\epsilon_1\cdot\epsilon_2\int_1^\infty ds\int_0^{1/s} d\nu\,\tilde{I}_0(\nu,s)\tilde{I}_1(\nu,s), 
\end{equation}
where 
\begin{eqnarray}
\tilde{I}_0(\nu,s) 
&:=& (8\pi^2\alpha^\prime  )^{-(p+1)/2}s^{(p-27)/2}e^{-2\pi r^2/s}\eta(is)^{-24} \nonumber \\ [2mm] 
& & \times \exp\left[ -2\alpha^\prime  k^2\log\left( -\frac1s\frac{\vartheta_{11}(\nu s,is)}{\eta(is)^3} \right) \right], 
   \label{integrand_gauge_UV-1} \\ [2mm] 
\tilde{I}_1(\nu,s) 
&:=& -\partial_\nu^2\log\left( -\frac1s\frac{\vartheta_{11}(\nu s,is)}{\eta(is)^3} \right). 
\end{eqnarray}
We have used the formula 
\begin{equation}
\pi\frac{\nu(t-\nu)}t+G_B(\nu,t)\ =\ \log\left( -\frac1s\frac{\vartheta_{11}(\nu s,is)}{\eta(is)^3} \right) 
\end{equation}
which is shown in appendix \ref{app_modular}. 
In the following, we investigate the expansions of ${\cal A}_{\rm IR}$ and ${\cal A}_{\rm UV}$ in detail. 

\vspace{5mm}

\subsubsection{The IR part ${\cal A}_{\rm IR}$}
\label{sec:PMT-ampIR}
\vspace{5mm}

We begin with the IR part ${\cal A}_{\rm IR}$. 
As for the effective potential, we can expand the integrand of ${\cal A}_{\rm IR}$ in terms of $q$, but with one caution. 
There is a new ingredient $G_B(\nu,t)$, a part of the worldsheet propagator, in the integrands (\ref{integrand_gauge_IR-1}) and (\ref{integrand_gauge_IR-2}). 
Most of the factors in the logarithm in (\ref{def_G_B}) can be set to 1 in the leading order approximation since their deviations from 1 are exponentially small all over the integration region $1\le t< \infty,\ 0\le \nu\le t$. 
However, there are two factors, $1-e^{-2\pi\nu}$ and $1-e^{-2\pi(t-\nu)}$, which may become close to 0 when $\nu$ approaches $0$ and $t$, respectively. 
Therefore, they must be retained in the approximation. 
Then $G_B(\nu,t)$ should be replaced with 
\begin{equation}
G_B^0(\nu,t)\ :=\ \log(1-e^{-2\pi\nu})+\log(1-e^{-2\pi(t-\nu)}). 
   \label{approx_G_B}
\end{equation}
Note that this is not negligible only when two vertex operators come close to each other. 
Otherwise $G_B^0(\nu,t)$ can be set to zero. 

Now we obtain the leading order approximation given as 
\begin{eqnarray}
{\cal A}_{\rm IR} 
&\sim& C\epsilon_1\cdot\epsilon_2\int_1^\infty dt\int_0^t d\nu\,(8\pi^2\alpha^\prime  t)^{-(p+1)/2}e^{-2\pi (r^2-1)t}\exp\left( -2\pi\alpha^\prime  k^2\frac{\nu(t-\nu)}t \right) \nonumber \\ [2mm] 
& & \hspace*{3cm}\times \exp\Bigl( -2\alpha^\prime  k^2G_B^0(\nu,t) \Bigr)\left( \frac{2\pi}t-\partial_\nu^2G_B^0(\nu,t) \right)
   \label{approx_gauge_IR}
\end{eqnarray}
where we have also replaced $\eta(it)^{-24}$ by $e^{2\pi t}$. 

We notice by setting $G_B^0(\nu,t)=0$ that (\ref{approx_gauge_IR}) includes a term 
\begin{equation}
C\epsilon_1\cdot\epsilon_2\int_1^\infty dt\int_0^t d\nu\,(8\pi^2\alpha^\prime  t)^{-(p+1)/2}e^{-2\pi (r^2-1)t}\exp\left( -2\pi\alpha^\prime  k^2\frac{\nu(t-\nu)}t \right)\frac{2\pi}t. 
\label{string-gauge2amplitude}
\end{equation}
This coincides with a gauge theory amplitude corresponding to the Feynman diagram shown in figure \ref{fig_Feynman_gauge}(a). 
The gauge theory result in the Schwinger parametrization is given in (\ref{C1gaugeamplitude})
 in appendix \ref{app_QFT_gauge}.
In the above (\ref{string-gauge2amplitude}), we have a UV cut-off at $t=1$. 
The UV cut-off scale introduced here corresponds to the string scale $\Lambda=1/\sqrt{\alpha^\prime}$. 
Comparing with the gauge theory amplitude, we can fix the overall constant $C$ of the amplitude to be 
\begin{equation}
C\ =\ \alpha^\prime  g_{\rm YM}^2.  
\end{equation}

\vspace{5mm}

In gauge theory, there is also a contact interaction corresponding to another Feynman diagram depicted in figure \ref{fig_Feynman_gauge}(b). 
This diagram does not depend on the external momentum $k^\mu$. 
The only possible term 
 in (\ref{approx_gauge_IR}), i.e., the only $k$-independent term except for the one in (\ref{string-gauge2amplitude}), is 
\begin{equation}
-\alpha^\prime  g_{\rm YM}^2\epsilon_1\cdot\epsilon_2\int_1^\infty dt\int_0^t d\nu\,(8\pi^2\alpha^\prime  t)^{-(p+1)/2}e^{-2\pi (r^2-1)t}\partial_\nu^2G_B^0(\nu,t). 
   \label{string_contact}
\end{equation}
This integral is divergent. 
This is because the integrand behaves as $\nu^{-2}$ and $(t-\nu)^{-2}$ at both ends of the integration region of $\nu$. 
We employ the zeta function regularization to deal with this integral. 
The result of the $\nu$-integration is then 
\begin{equation}
\alpha^\prime  g_{\rm YM}^2\epsilon_1\cdot\epsilon_2\int_1^\infty dt\,(8\pi^2\alpha^\prime  t)^{-(p+1)/2}e^{-2\pi (r^2-1)t}\cdot(-2\pi). 
   \label{gauge_contact}
\end{equation}
See appendix \ref{app_leading} for details. 
This exactly coincides with the gauge theory amplitude for the contact interaction in figure \ref{fig_Feynman_gauge}(b). 
This coincidence shows that the zeta function regularization is consistent with the gauge invariance of the worldvolume effective theory of D-branes. 

\vspace{5mm}

The gauge theory amplitudes have been exhausted, but there are still many terms in ${\cal A}_{\rm IR}$ besides the field theory result. 
No other terms of order ${\cal O}(k^0)$ exist, as expected from the gauge invariance. 
At the order ${\cal O}(k^2)$, the integral (\ref{approx_gauge_IR}) has terms 
\begin{eqnarray}
& & \alpha^\prime  g_{\rm YM}^2\epsilon_1\cdot\epsilon_2\,\alpha^\prime  k^2\int_1^\infty dt\int_0^t d\nu\,(8\pi^2\alpha^\prime  t)^{-(p+1)/2}e^{-2\pi (r^2-1)t} \nonumber \\ [2mm] 
& & \hspace*{2cm}\times\left[ -\frac{2\pi}t\cdot 2G_B^0(\nu,t)+\left( 2\pi\frac{\nu(t-\nu)}t+2G_B^0(\nu,t) \right)\partial_\nu^2G_B^0(\nu,t) \right]. 
\end{eqnarray}
Since $\alpha^\prime  $ is essentially the only dimensionful parameter, all these terms give contributions of the same order with the field theory amplitudes. 
After performing the $\nu$-integration, we obtain 
\begin{equation}
\alpha^\prime  g_{\rm YM}^2\epsilon_1\cdot\epsilon_2\,\alpha^\prime  k^2\int_1^\infty dt\,(8\pi^2\alpha^\prime  t)^{-(p+1)/2}e^{-2\pi(r^2-1)t}\left( \frac{4\pi^2}{3t}+4\pi \right). 
   \label{stringy_open}
\end{equation}
See appendix \ref{app_leading} for details.  
It is interpreted as an IR remnant of the stringy effects. 
In the next section, we will show that the above term becomes negligible when we shift the division point from $t=1$ to a larger value, corresponding to a smaller UV cut-off. 

These extra contributions come from terms in ${\cal A}_{\rm IR}$ including $G_B^0(\nu,t)$. 
In the beginning of this section \ref{sec:PMT-ampIR}, we noticed that $G_B^0(\nu,t)$ is non-negligible only when the vertex operators are about to collide. 
In string theory, when two or more vertex operators come close together, then the situation is equivalent by the conformal invariance to the development of a long and thin strip connecting two worldsheets. 
The closed string analog of this phenomenon is depicted in figure \ref{fig_Riemann_surface}(b). 
This strip indicates the propagation of various open string modes including massive ones. 
Therefore, the corners $\nu\sim 0,t$ of the integration region are the regions where stringy effects become relevant. 
Thus the extra terms (\ref{stringy_open}) can be interpreted as remnants of the stringy effects. 

It is interesting to observe that in the calculation of the contact interaction (\ref{gauge_contact}), 
on the other hand, we obtained only the gauge theory amplitude and no stringy corrections. 
Indeed, this is required by the gauge invariance. 
This result also implies that, at least for the string amplitude considered in this section, all the possible contributions from the propagation of open string modes are automatically subtracted appropriately by the zeta function regularization. 
In other words, only 1PI diagrams are retained by the zeta function regularization. 
This is an advantage of this regularization scheme. 

\vspace{5mm}

\subsubsection{The UV part ${\cal A}_{\rm UV}$} \label{sec:PMT_UV}

\vspace{5mm}

The UV part  ${\cal A}_{\rm UV}$ can be interpreted as  purely stringy corrections to the amplitude from the  
effective field theory point of view on the world volume. 
In order to expand the integrand $\tilde{I}_0(\nu,s)\tilde{I}_1(\nu,s)$ of ${\cal A}_{\rm UV}$, we use the formula 
\begin{equation}
\log\left( -\frac1s\frac{\vartheta_{11}(\nu s,is)}{\eta(is)^3} \right)\ =\ -\sum_{n=1}^\infty\frac1n\frac{1+\tilde{q}^n}{1-\tilde{q}^n}\cos(2\pi n\nu s)-\log\left( s\prod_{n=1}^\infty(1-\tilde{q}^n)^2 \right). 
   \label{Fourier_expansion}
\end{equation}
This is shown in appendix \ref{app_modular}. 
Then, the leading order approximation to ${\cal A}_{\rm UV}$ is given as 
\begin{eqnarray}
{\cal A}_{\rm UV} 
&\sim& -\alpha^\prime  g_{\rm YM}^2\epsilon_1\cdot\epsilon_2\int_1^\infty ds\int_0^{1/s}d\nu\,(8\pi^2\alpha^\prime  )^{-(p+1)/2}s^{(p-27)/2}e^{-2\pi r^2/s} 
   \label{approx_gauge_UV} \\
& & \times 24\exp\left[ 2\alpha^\prime  k^2\left( \sum_{n=1}^\infty\frac1n\cos(2\pi n\nu s)-\log s \right) \right]\cdot(2\pi s)^2\sum_{n=1}^\infty n\cos(2\pi n\nu s). \nonumber 
\end{eqnarray}
Here we have expanded 
\begin{equation}
\eta(is)^{-24}\ =\ e^{2\pi s}+24+{\cal O}(e^{-2\pi s}), 
\label{eta-expansion}
\end{equation}
and eliminated by hand the term $e^{2\pi s}$, corresponding to the closed string tachyon. 
As explained in the introduction, the UV part ${\cal A}_{\rm UV}$ is expected to give corrections coming from high energy modes to the low-energy effective action, a part of which is contained in ${\cal A}_{\rm IR}$ as we have just observed. 
In order to obtain explicit expressions, we expand (\ref{approx_gauge_UV}) in terms of the external momentum, and calculate those coefficients. 

The term independent of the external momentum is 
\begin{eqnarray}
& & -24\alpha^\prime  g_{\rm YM}^2\epsilon_1\cdot\epsilon_2\int_1^\infty ds\,(8\pi^2\alpha^\prime  )^{-(p+1)/2}s^{(p-27)/2}e^{-2\pi r^2/s} \nonumber \\
& & \times \int_0^{1/s}d\nu\,(2\pi s)^2\sum_{n=1}^\infty n\cos(2\pi n\nu s). 
\end{eqnarray}
Integrating term by term, we find that this contibution vanishes. 
This is consistent with the gauge invariance since a term like this gives a mass shift to the gauge boson if it is non-vanishing. 

At order ${\cal O}(k^2)$, we have 
\begin{eqnarray}
& & -48\alpha^\prime  g_{\rm YM}^2\epsilon_1\cdot\epsilon_2\,\alpha^\prime  k^2\int_1^\infty ds\,(8\pi^2\alpha^\prime  )^{-(p+1)/2}s^{(p-27)/2}e^{-2\pi r^2/s} \nonumber \\
& & \times \int_0^{1/s}d\nu \left( \sum_{n=1}^\infty\frac1n\cos(2\pi n\nu s)-\log s \right)\cdot(2\pi s)^2\sum_{n=1}^\infty n\cos(2\pi n\nu s). 
\end{eqnarray}
The $\nu$-integration can be easily performed, resulting in 
\begin{equation}
-48\pi^2\alpha^\prime  g_{\rm YM}^2\epsilon_1\cdot\epsilon_2\,\alpha^\prime  k^2\int_1^\infty ds\,(8\pi^2\alpha^\prime  )^{-(p+1)/2}s^{(p-25)/2}e^{-2\pi r^2/s}. 
   \label{bosonic_UV_wf}
\end{equation}
This should be regarded as a sum of threshold corrections to the wave function renormalization due to an infinite number of massive open string states. 
Note that the overall coefficient of this term would depend on how we subtract the closed string tachyon contribution in the present setup. 
Therefore, the numerical value of this term could not be taken at its face value. 
In section \ref{sec:typeII}, we will investigate other setups without the closed string tachyon, in which we can discuss the correction terms for the wave function renormalization. 

\vspace{5mm}

\begin{figure}
\begin{center}
\includegraphics{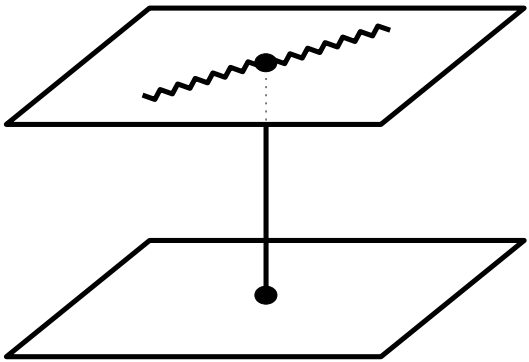}
\end{center}
\caption{The diagram for the exchange of closed string states corresponding to the wave function renormalization of the gauge bosons. 
The wavy lines represent the gauge bosons, and the solid line represents closed string states. }
   \label{fig_closed_channel}
\end{figure}

Due to the open-closed duality, we expect that the amplitude (\ref{bosonic_UV_wf}) can be reproduced as the amplitudes for the exchange of massless closed string modes between the D-branes, with possible stringy corrections. 
The diagrams for the exchanges are like the one depicted in figure \ref{fig_closed_channel}. 
Note that this is a {\it tree} diagram for the closed string. 
Therefore, there is no parameter like $\nu$ which should be integrated. 
Fortunately, we can perform the $\nu$-integration in (\ref{approx_gauge_UV}) at least order by order in $k^2$, as shown above \cite{DiVecchia:1996uq}. 
Therefore, the comparison of (\ref{bosonic_UV_wf}) with the gravity amplitudes could be easily performed, as long as the off-shell couplings of massless closed string modes to the D-brane are clarified. 
We can use it to extract the leading behavior of possible stringy corrections. 
It will be discussed further in sections \ref{sec:renormUV} and  \ref{sec:typeIIUV}. 

\vspace{5mm}

\subsubsection{Comparison with the case of effective potential}

\vspace{5mm}

In the calculation of the effective potential using the partial modular transformation, reviewed in section \ref{subsec_effective_potential}, we can obtain a good approximation of the potential for arbitrary distance $r$ between D-branes by using only gauge theory and gravity at low-energy, and its corrections are largely suppressed by a factor $e^{-2\pi}$.
On the other hand, as we saw in this section \ref{sec:PMTstringamplitude}, 
stringy corrections become sizable for the scattering amplitudes when the vertices are close to each other. 
Why are we able to avoid receiving sizable stringy corrections for the calculation of the effective potential although it is related to that of an amplitude? 

The reason is the following. 
In the calculation of the effective potential in quantum field theory, we replace fields, whose potential is of interest, with their expectation values. 
The corresponding calculation in string theory should be the one in which vertex operators for the fields are replaced with certain c-numbers. 
Obviously nothing will happen when those c-numbers collide. 
Therefore, any kinds of stringy corrections should be absent. 

\vspace{1cm}

\section{Partial modular transformation and Wilsonian renormalization} \label{sec:renorm}

\vspace{5mm}

In the previous section, we found that the partial modular transformation allows us to interpret string amplitudes ${\cal A}={\cal A}_{\rm IR}+{\cal A}_{\rm UV}$ approximately as a sum of gauge theory amplitudes and gravity amplitudes, with some stringy corrections of the same order. 
The choice of the division point $t=1$ is required such that both ${\cal A}_{\rm IR}$ and ${\cal A}_{\rm UV}$ can be expanded simultaneously in terms of a small parameter $e^{-2\pi}$. 
For this choice, we observed that threshold corrections due to all massive open string modes can be described by a few terms at each order of the derivative expansion. 
It is obvious that sub-leading corrections can be systematically included as well. 

Recall that a UV cut-off of the string scale is introduced in the gauge theory amplitudes obtained from ${\cal A}_{\rm IR}$. 
Suppose that the division point is shifted to $t=\lambda^{-2}$. 
For this choice, the UV cut-off in the mass unit becomes 
\begin{equation}
\Lambda\ :=\ \frac\lambda{\sqrt{\alpha^\prime}}. 
\end{equation}
If we choose $\lambda$ to be small, then $\Lambda$ becomes much smaller than the string scale. 
This implies that the stringy corrections in ${\cal A}_{\rm IR}$ discussed in the previous section becomes negligible compared to the effective field theory amplitudes, as long as the external momentum is smaller than $\Lambda$. 
This is simply because the stringy corrections contain a power of $\alpha^\prime \Lambda^2=\lambda^2$. 
We will show this explicitly in the following. 
The UV part ${\cal A}_{\rm UV}$ then contains all the stringy corrections. 
This is reminiscent of the Wilsonian renormalization: the UV part of the amplitude can be regarded as contributions due to effective vertices obtained by integrating out high-energy modes. 

Note that taking small $\lambda$ is effectively equivalent to the {\it zero slope limit} for ${\cal A}_{\rm IR}$, and we will use this term in the following. 
However, it is important to keep in mind that we always treat $\alpha^\prime$ as a fixed constant, since otherwise the whole amplitude is reduced to the effective field theory amplitude without stringy corrections. 

\vspace{5mm}

In this section, we consider a 2-point amplitude in bosonic string theory whose external states are open string tachyons, instead of gauge bosons discussed in the previous section. 
We choose this setup because there is no constraint of gauge invariance, and therefore, we can discuss stringy corrections to the mass. 
The string diagram is of the type in figure \ref{fig_1-loop_planar}. 
The amplitude is given as 
\begin{equation}
{\cal A}\ =\ C'\int_0^\infty dt\int_0^td\nu\,I_0(\nu,t), 
\end{equation}
where $I_0(t)$ is given in (\ref{integrand_gauge_IR-1}). 
We simply ignore the other string diagrams of the types in figure \ref{fig_1-loop_irrelevant} since their IR divergences are difficult to regularize. 
Recall that we can take the external momentum $k^\mu$ to be off-shell, so that $k^2$ does not need to be of the string scale. 

We apply the partial modular transformation to this amplitude with the division point $t=\lambda^{-2}$. 
As a result, we obtain 
\begin{equation}
{\cal A}\ =\ {\cal A}_{\rm IR}+{\cal A}_{\rm UV}, 
\end{equation}
where 
\begin{equation}
{\cal A}_{\rm IR}\ =\ C'\int_{\lambda^{-2}}^\infty dt\int_0^td\nu\,I_0(\nu,t), \hspace{1cm} {\cal A}_{\rm UV}\ =\ C'\int_{\lambda^2}^\infty ds\int_0^{1/s}d\nu\,\tilde{I}_0(\nu,s). 
\end{equation}
The integrand $\tilde{I}_0(s)$ is given in (\ref{integrand_gauge_UV-1}). 
In the following, we consider the limiting case of $\lambda\ll1$.  

\vspace{5mm}

\subsection{The IR part ${\cal A}_{\rm IR}$}

\vspace{5mm}

We investigate the relation between ${\cal A}_{\rm IR}$ and the corresponding field theory amplitude given in appendix \ref{app_QFT_tachyon}. 
Since $q\le e^{-2\pi/\lambda^2}$ is negligibly small, we can approximate ${\cal A}_{\rm IR}$ as 
\begin{eqnarray}
{\cal A}_{\rm IR} 
&\sim& C'\int_{\lambda^{-2}}^\infty dt\int_0^t d\nu\,(8\pi^2\alpha^\prime t)^{-(p+1)/2}e^{-2\pi (r^2-1)t} \nonumber \\ [2mm] 
& & \hspace*{2cm}\times\exp\left( -2\pi \alpha^\prime k^2\frac{\nu(t-\nu)}t-2\alpha^\prime k^2G_B^0(\nu,t) \right), 
   \label{approx_tachyon_IR}
\end{eqnarray}
where $G_B^0(\nu,t)$ in (\ref{approx_G_B}) is an approximation of $G_B(\nu,t)$.  
It is easy to notice that the part 
\begin{equation}
C'\int_{\lambda^{-2}}^\infty dt\int_0^t d\nu\,(8\pi^2\alpha^\prime t)^{-(p+1)/2}e^{-2\pi (r^2-1)t}\exp\left( -2\pi\alpha^\prime k^2\frac{\nu(t-\nu)}t \right)
\end{equation}
coincides with the field theory amplitude (\ref{app_amplitude_tachyon}), provided 
\begin{equation}
C'\ =\ (2\pi)^2g^2(\alpha^\prime)^2. 
\end{equation}
Here $g$ is the coupling constant for the vertex $T_0|T|^2$, where $T_0$ is a real tachyon field coming from an open string living on one D-brane, and $T$ is a complex tachyon field coming from the stretched string between the D-branes. 

The other terms in ${\cal A}_{\rm IR}$ give stringy corrections. 
There is no correction at order ${\cal O}(k^0)$. 
At the order ${\cal O}(k^2)$, we have 
\begin{equation}
(2\pi)^2g^2(\alpha^\prime)^2\int_{\lambda^{-2}}^\infty dt\int_0^td\nu\,(8\pi^2\alpha^\prime t)^{-(p+1)/2}e^{-2\pi(r^2-1)t}\Bigl( -2\alpha^\prime k^2G_B^0(\nu,t) \Bigr). 
\end{equation}
The $\nu$-integration can be performed as shown in appendix \ref{app_leading}. 
As a result, and after a rescaling of $t$, we obtain 
\begin{equation}
\frac\pi3(2\pi)^2g^2(\alpha^\prime)^2k^2\int_{\Lambda^{-2}}^\infty d\tau\,(8\pi^2\tau)^{-(p+1)/2}e^{-2\pi m^2\tau}, 
\end{equation}
up to exponentially small terms. 
Note that the integration variable $\tau:=\alpha^\prime t$ has mass dimension $-2$. 
This correction has extra $\alpha^\prime$ dependence compared to the field theory amplitude (\ref{app_amplitude_tachyon1}). 
By rewriting 
\begin{equation}
(\alpha^\prime)^2k^2\ =\ \lambda^4\frac{k^2}{\Lambda^4}, 
\end{equation}
we see that this stringy correction is negligible for small $\lambda$, as long as $k^2$ is smaller than the cut-off scale. 
We can even take a limit in which $\lambda$ goes to zero while $\Lambda$ is fixed. 
This implies we also take $\alpha^\prime$ to zero, and therefore, this is nothing but the zero slope limit. 
We see that the above correction vanishes in the zero slope limit, provided that $m^2=(r^2-1)/\alpha^\prime$ is also kept fixed. 

This is also valid for other corrections. 
It is easily recognized by rescaling the integration variable $t$ in (\ref{approx_tachyon_IR}) as 
\begin{eqnarray}
{\cal A}_{\rm IR} 
&\sim& (2\pi)^2g^2\int_{\Lambda^{-2}}^\infty d\tau\int_0^\tau d\tau_2\,(8\pi^2\tau)^{-(p+1)/2}e^{-2\pi m^2\tau} \nonumber \\ [2mm] 
& & \hspace*{2cm}\times\exp\left( -2\pi k^2\frac{\tau_2(\tau-\tau_2)}\tau-2\alpha^\prime k^2G_B^0(\tau_2/\alpha^\prime,\tau/\alpha^\prime) \right). 
\end{eqnarray}
Now the $\alpha^\prime$-dependence is accumulated in the second term in the exponential. 
We find that the $\tau_2$-integral in the presence of $G_B^0(\tau_2/\alpha^\prime,\tau/\alpha)$ can be well estimated by an integral around $\tau_2=0$ and $\tau_2=\tau$  (see \ref{approx_G_B}) as 
\begin{equation}
\int_0^\tau d\tau_2\,(G_B^0)^n\ \sim\ \int_0^{\alpha^\prime}d\tau_2\,(G_B^0)^n+\int_{\tau-\alpha^\prime}^\tau d\tau_2\,(G_B^0)^n\ =\ {\cal O}(\alpha^\prime). 
\end{equation}
Therefore, the stringy corrections are always higher orders in $\alpha^\prime$, or equivalently, higher orders in $\lambda^2/\Lambda^2$, and they vanish in the zero slope limit. 

\vspace{5mm}

It is instructive to consider the zero slope limit of the gauge boson amplitude investigated in section \ref{sec:PMT}. 
The amplitude in rescaled variables is 
\begin{eqnarray}
& & g_{\rm YM}^2\epsilon_1\cdot\epsilon_2\int_{\Lambda^{-2}}^\infty d\tau\int_0^\tau d\tau_2\,(8\pi^2\tau)^{-(p+1)/2}e^{-2\pi m^2\tau}
\\ [2mm] & & 
\times
\exp\left( -2\pi k^2\frac{\tau_2(\tau-\tau_2)}\tau-2\alpha^\prime k^2G_B^0(\tau_2/\alpha^\prime,\tau/\alpha^\prime) \right) 
\left( \frac{2\pi}\tau-\frac1{\alpha^\prime}\partial_x^2G_B^0(x,\tau/\alpha^\prime)\Big|_{x=\tau_2/\alpha^\prime} \right).
\nonumber
\end{eqnarray}
As for the tachyon amplitude, the factor $2\alpha^\prime k^2G_B^0(\tau_2/\alpha^\prime,\tau/\alpha^\prime)$ raises the order of $\alpha^\prime$. 
The difference is the presence of the inverse power $(\alpha^\prime)^{-1}$ in front of $\partial_x^2G_B^0(x,\tau/\alpha^\prime)$. 
This enables the amplitude (\ref{app_amplitude_contact}) for the contact interaction to survive in the zero slope limit. 

\vspace{5mm}

\subsection{The UV part ${\cal A}_{\rm UV}$}
\label{sec:renormUV}
\vspace{5mm}

We have observed that ${\cal A}_{\rm IR}$ approaches the amplitude for the low-energy effective theory when $\lambda$ is small. 
The stringy corrections, which are of the same order with the field theory amplitudes for $\lambda=1$, are now negligible. 
Then, all the stringy corrections are effectively contained in the UV part ${\cal A}_{\rm UV}$ of the amplitude. 
This is given as 
\begin{eqnarray}
{\cal A}_{\rm UV} 
&=& (2\pi)^2g^2(\alpha^\prime)^2\int_{\lambda^2}^\infty ds\int_0^{1/s}d\nu\,(8\pi^2\alpha^\prime)^{-(p+1)/2}s^{(p-27)/2}e^{-2\pi r^2/s}\eta(is)^{-24} \nonumber \\ [2mm] 
& & \hspace*{3cm}\times \exp\left[ -2\alpha^\prime k^2\log\left( -\frac1s\frac{\vartheta_{11}(\nu s,is)}{\eta(is)^3} \right) \right]. 
\end{eqnarray}
Now $\tilde{q}\le e^{-2\pi\lambda^2}$ can become close to 1 when $\lambda$ is small. 
Therefore, we have to keep all higher powers of $\tilde{q}$ in the integrand. 

Recall that the logarithm in the exponent can be rewritten as (\ref{Fourier_expansion}). 
Then, the $\nu$-integration can be performed exactly for each term in the $k^2$-expansion of the integrand as \cite{DiVecchia:1996uq}
\begin{eqnarray}
& & \int_0^{1/s}d\nu\,\exp\left[ -2\alpha^\prime k^2\log\left( -\frac1s\frac{\vartheta_{11}(\nu s,is)}{\eta(is)^3} \right) \right] \nonumber \\ [2mm] 
&=& \frac1s+\frac{2\alpha^\prime k^2}s\log\left( s\prod_{m=1}^\infty(1-\tilde{q}^m)^2 \right)+{\cal O}(k^4). 
\end{eqnarray}
This is an advantage for the interpretation of ${\cal A}_{\rm UV}$ as corrections of the low-energy amplitude coming from integrating out high-energy modes. 
The integration variable $\nu$ indicates the distance  of the vertex operators on the worldsheet, which has no interpretation for local operators in the low-energy effective theory. 
In the absence of $\nu$, we can simply regard the coefficients in the $k^2$-expansion of ${\cal A}_{\rm UV}$ as coefficients of local operators in the Wilsonian effective action which are suppose to provide the stringy corrections. 

Up to this point, however, the calculations for ${\cal A}_{\rm UV}$ is just formal ones since the integrals are divergent due to the presence of the closed string tachyon in bosonic string theory. 
This is not easily eliminated like the open string tachyon circulating in the loop. 
In the following, as in section \ref{sec:PMT_UV}, we simply get rid of the closed string tachyon by hand and obtain finite quantities. 
This is just for an illustrative purpose, and the physical relevance of the resulting finite quantities is obscure. 
In section \ref{sec:typeII}, we analyze amplitudes in the absence of closed string tachyons. 
In these setups, we can discuss the physical significance of these stringy corrections, in particular, the sign and the magnitude. 

\vspace{5mm}

(i) {\it Mass shifts} 

\vspace{5mm}

First, we discuss the mass shift obtained from ${\cal A}_{\rm UV}$. 
It is given as 
\begin{equation}
-(2\pi)^2g^2(\alpha^\prime)^2(8\pi^2\alpha^\prime)^{-(p+1)/2}\int_{\lambda^2}^\infty ds\,s^{(p-29)/2}e^{-2\pi r^2/s}\Bigl( \eta(is)^{-24}-e^{2\pi s} \Bigr). 
   \label{mass_shift_bosonic}
\end{equation}
Note that the contribution of the closed string tachyon is subtracted, as explained above (see (\ref{eta-expansion})). 
The integral is now finite even in the small $\lambda$ limit. 
It is convenient to rewrite the overall factor as $\xi_0/2\alpha^\prime$ where 
\begin{equation}
\xi_0\ :=\ (8\pi^2)^{(1-p)/2}g^2 ({\alpha^\prime})^{3-(p+1)/2}
\end{equation}
is a dimensionless quantity. 
Then, it becomes apparent that the mass dimension of (\ref{mass_shift_bosonic}) is 2. 

Various integral formulas for mass shifts can be found in the literature. 
In our setup, we have a parameter $r$, the (dimensionless) distance between the D-branes. 
By using this ingredient, we can further investigate the mass shift as follows. 
We expand $\eta(is)^{-24}$ as 
\begin{equation}
\eta(is)^{-24}\ =\ e^{2\pi s}+\sum_{n=0}^\infty d_n\tilde{q}^n, 
\end{equation}
and integrate term by term. 
Then, we obtain 
\begin{eqnarray}
& & \int_{\lambda^2}^\infty ds\,s^{(p-29)/2}e^{-2\pi r^2/s}\Bigl( \eta(is)^{-24}-e^{2\pi s} \Bigr) \nonumber \\
&\to& 24\,\Gamma({\textstyle \frac{27-p}2})(2\pi r^2)^{(p-27)/2}+\sum_{n=1}^\infty \frac{d_n}{2\pi n}(8\pi^2n)^{(29-p)/2}g_{29-p}(4\pi\sqrt{n}r), 
\end{eqnarray}
where the limit $\lambda\to0$ has been taken. 
The functions $g_D(x)$ are defined in appendix \ref{app_bulk}. 
For large $x$, they behave as 
\begin{equation}
g_D(x)\ \sim\ \frac12(2\pi)^{(1-D)/2}x^{(1-D)/2}e^{-x}. 
\end{equation}
Therefore, this gives the expansion of the mass shift for large $r$. 

We can understand the dependence of the mass shift on $r$ as follows. 
Recall that the amplitude (\ref{mass_shift_bosonic}) is written in the closed string channel. 
Then, each term given in the $\tilde{q}$-expansion of $\eta(is)^{-24}$ gives the amplitude for the exchange of closed string modes. 
The term proportional to $r^{p-27}$ comes from the exchange of massless closed string states. 
This looks like the Newtonian potential, but the power is different due to the presence of the external states. 
On the other hand, the term proportional to $g_{29-p}(4\pi\sqrt{n}r)$ comes from the exchanges of the $n$-th massive closed string states, resulting in the functions like the Yukawa potential. 

One may be interested in the sign of the mass shift. 
It can be read off, in principle, by comparing the above expression with the field theory amplitude. 
In this bosonic string setup, however, the sign would be ambiguous since the subtraction scheme for the closed string tachyon might change the overall coefficient. 
In section \ref{sec:typeII}, we can safely discuss the sign of the stringy mass shift in superstring setups. 

\vspace{5mm}
(ii) {\it A possibility to obtain a hierarchical mass spectrum} 

\vspace{5mm}

Although the overall coefficient of the mass shift (\ref{mass_shift_bosonic}) would be ambiguous, there is one lesson we can learn which might be  important for a phenomenological reason. 
Suppose that, in a suitable setup, the term proportional to a power of $r$ is absent. 
Then, the leading contribution to the mass shift is given by the Yukawa-type function. 
This is an exponentially decreasing function of $r$, so even when $r$ is of order 5, that is, the distance between the D-branes is about 5 times the string scale, the stringy correction to the mass is far smaller than the string scale, even after taking into account all massive open string modes. 
It is then natural to expect that this would give us a mechanism to realize a hierarchical mass spectrum in the low-energy field theory on a D-brane system. 
A similar phenomenon is observed in the context of closed string theory in \cite{Itoyama:2019yst,Itoyama:2020ifw}. 

Recall that the term proportional to a power of $r$ is given by the gravity amplitude depicted in figure \ref{fig_closed_channel}. 
This can be evaluated simply by using the low-energy gravity theory, provided that the couplings of the massless closed string states to D-branes, including stringy corrections to the DBI action, are obtained. 
Therefore, we do not need to calculate 1-loop string amplitudes in order to check whether the exponentially small mass corrections can be obtained for a given D-brane system. 
The couplings of massless closed string modes to D-branes were investigated in \cite{Hashimoto:1996kf} in the context of the Hawking radiation from D-branes. 

Even if the massless closed string exchanges do not cancel, the above structure of the mass shift tells us another interesting fact: the mass shift including threshold corrections from all massive open string modes can be calculated by evaluating the gravity amplitudes like the one in figure \ref{fig_closed_channel}. 

\vspace{5mm}

(iii) {\it Wave function renormalizations} 

\vspace{5mm}

We can also discuss the wave function renormalization due to stringy corrections. 
In fact, this is crucial for obtaining the physical mass from the mass shift determined above. 
By the reasons explained above, we will postpone the discussion of their physical meaning, and just present the resulting expression 
\begin{equation}
\xi_0k^2\int_{\lambda^2}^\infty ds\,s^{(p-29)/2}e^{-2\pi r^2/s}\Bigl( \eta(is)-e^{2\pi s} \Bigr)\log\left( s\prod_{m=1}^\infty(1-\tilde{q}^m)^2 \right). 
\end{equation}
The leading order contribution to this integral in the small $\lambda$ limit is 
\begin{equation}
\xi_0k^2\Gamma({\textstyle \frac{27-p}2})\Bigl( \log(2\pi r^2)-\psi({\textstyle \frac{27-p}2}) \Bigr)(2\pi r^2)^{(p-27)/2}, 
\end{equation}
where $\psi(x)$ is the digamma function. 

\vspace{1cm}

\section{Mass shifts in Type II string theory} \label{sec:typeII}

\vspace{5mm}

In the previous section, we encountered the problem of divergence due to the presence of a closed string tachyon. 
This problem prevented us from extracting definite physical information on mass shifts and wave function renormalizations. 
In this section, we consider similar amplitudes in Type II string theory where a closed string tachyon is absent in the bulk. 
We investigate mass shifts of open string states on a D-brane, which are massless at tree level. 
We will show that the mass shifts can be negative, and as a result, we may obtain a Higgs-like field. 
The mass shift can be exponentially small in a particular setup. 

\vspace{5mm}

\subsection{Green-Schwarz formalism} \label{sec:typeII_GS}

\vspace{5mm}

We need to calculate 2-point amplitudes in Type II string theory. 
The Green-Schwarz formalism in the light cone gauge \cite{Green:1987sp,Green:1987mn} is  convenient for this purpose. 
To fix notations, we write down the worldsheet action 
\begin{equation}
S\ =\ \int d^2\sigma\left[ -\frac1{4\pi\alpha^\prime}\partial_\alpha X^i\partial^\alpha X^i+\frac i{2\pi}S^a(\partial_\tau+\partial_\sigma)S^a+\frac i{2\pi}\tilde{S}^a(\partial_\tau-\partial_\sigma)\tilde{S}^a \right] 
   \label{GS_action}
\end{equation}
for Type IIB string theory. 
For Type IIA, the fermionic fields $\tilde{S}^a$ are replaced with $\tilde{S}^{\dot{a}}$. 
The indices $i,a,\dot{a}$ all run from 1 to 8, and they label the vector, spinor, and conjugate spinor representations of ${\rm Spin}(8)$, respectively. 
For practical purposes, this gauge-fixed action can be regarded as a definition of the worldsheet theory. 
As long as the kinematics of the external states is simple, we can recover covariant forms of the amplitudes from the expressions obtained in the Green-Schwarz formalism. 
As explained in appendix \ref{app_amplitude}, we can obtain 2-point amplitudes of interest partly by recycling the amplitudes in bosonic string theory we have discussed so far. 

\vspace{5mm}

As before, we consider the scattering of two open string states on a D-brane, as in figure \ref{fig_1-loop_planar}. 
In order to calculate the amplitude, we need to quantize the open string stretched between the D-branes. 
This then requires us to specify the corresponding boundary condition for the open string. 
The boundary conditions for $X^i$ are obvious from the configuration of the D-branes. 
In the Green-Schwarz formalism, the boundary conditions for $S^a$ and $\tilde{S}^a$ can be read off from the space-time supersymmetry conserved by the D-branes as follows. 
A D$p$-brane conserves the supercharges of the form \cite{Polchinski:1998rr} 
\begin{equation}
Q_\alpha+(\beta^\perp\tilde{Q})_\alpha, 
\end{equation}
where $\beta^\perp$ is a product of the Gamma matrices corresponding to the transverse directions to the D$p$-brane. 
In the Green-Schwarz formalism, the supercharges can be constructed from the worldsheet fields. 
The above combinations of the supercharges are conserved in the worldsheet theory, if we require that the corresponding supercurrents vanish at the endpoint of the open string attached to the D$p$-brane. 
Another condition for the supercurrents is obtained from the other endpoint. 
These two conditions are combined to give the correct boundary condition for the open string in the Green-Schwarz formalism. 
See appendix \ref{boundary_condition_GS} for more details. 

As in bosonic string theory, the 2-point 1-loop amplitude is given schematically as 
\begin{equation}
{\cal A}\ =\ (2\pi)^2\int_0^\infty dt_1\int_0^\infty dt_2\,{\rm STr}\Bigl( e^{-2\pi t_1(L_0-1/2)}V_1e^{-2\pi t_2(L_0-1/2)}V_2 \Bigr), 
\end{equation}
where we take the supertrace because of the presence of space-time fermions. 
The vertex operators $V_s$ ($s=1,2$) for gauge bosons on a D$9$-brane are given in \cite{Green:1987sp} as
\begin{equation}
V_s\ =\ c\,\epsilon_{s,i}\left( \partial_\tau X^i-\frac{\alpha^\prime}2\gamma^{ij}_{ab}k_{s,j}S^aS^b \right)e^{ik_{s,i}X^i}. 
   \label{vertex_typeII_Neumann}
\end{equation}
The constant $c$ will be chosen by comparing with the corresponding field theory amplitudes. 
The operators for D$p$-branes with $p<9$ can be obtained by performing the T-duality transformation \cite{Hashimoto:1996kf}. 
For bosonic fields $X^i$, this amounts to replacing $\partial_\tau X^i$ with $\partial_\sigma X^i$ for the directions transverse to the D$p$-brane. 
The T-duality transformation for ${\rm Spin}(9,1)$ spinor fields $\theta$ in the Green-Schwarz formalism, 
from which the ${\rm Spin}(8)$ spinor field $S^a$ is obtained, is given in \cite{Kulik:2000nr} as
\begin{equation}
\theta\ \to\ \Gamma^i\theta
\end{equation}
for each direction along which T-duality is taken. 
This transformation, written in terms of $S^a$, keeps the fermion bilinear term in $V_s$ intact for the gauge bosons, while for the transverse scalars it changes the sign. 
See appendix \ref{app_amplitude} for more details. 
As a result, the vertex operators for transverse scalars are 
\begin{equation}
V_s\ =\ c\,\zeta_{s,I}\left( \partial_\sigma X^I+\frac{\alpha^\prime}2\gamma^{Il}_{ab}k_{s,l}S^aS^b \right)e^{ik_{s,l}X^l},  
\end{equation}
where $I=1,\cdots,9-p$ labels the transverse directions, and $l=10-p,\cdots,8$ labels the parallel directions, except for the light cone directions, of the D$p$-brane. 
The vectors $\zeta_{s,I}$ specify transverse directions corresponding to the external scalars. 

\vspace{5mm}

\subsection{D$p$-$\overline{{\rm D}p}$ system}
\label{sec:typeIIppbar}
\vspace{5mm}

In this subsection, we consider a D$p$-brane and a $\overline{{\rm D}p}$-brane parallel to each other. 
The 2-point amplitude of our interest is for the massless transverse scalars on the D$p$-brane. 
In absence of the $\overline{{\rm D}p}$-brane, the scalars are the Nambu-Goldstone bosons for the translational symmetry. 
In our setup, the translational symmetry for the D$p$-brane is broken by the $\overline{{\rm D}p}$-brane, and therefore, the scalars may acquire masses by quantum corrections. 

There are string 1-loop diagrams contributing to this amplitude. 
The diagrams, which are of the types in figure \ref{fig_1-loop_irrelevant} but the external states are scalars, give vanishing contributions since the $\overline{{\rm D}p}$-brane is not involved in this diagram, and such supersymmetric 2-point amplitudes are known to vanish due to fermion zero modes \cite{Green:1987mn}. 
Therefore, we focus our attention on the type of the  diagram  in figure \ref{fig_1-loop_planar}. 
As explained in appendix \ref{boundary_condition_GS}, it turns out that the fermionic fields $S^a$ of the open string stretched between the D$p$-brane and the $\overline{{\rm D}p}$-brane obey the anti-periodic (Neveu-Schwarz (NS) type) boundary condition, as opposed to the periodic (Ramond (R) type) boundary condition obeyed by the open string on D$p$-branes. 
Because of this, the ground state of the open string gives a tachyon. 
In order to remove the IR divergence due to this open string tachyon, 
we  keep the distance, denoted by $r$, between the D-branes sufficiently large. 

The 2-point amplitude of the massless transverse scalars in this setup is given as 
\begin{eqnarray}
   {\cal A} \ &=& \ {\cal A}^1  +  {\cal A}^2  \nonumber \\ [2mm] 
{\cal A}^1 &=&
\ C^{\prime\prime}\int_0^\infty dt\int_0^t d\nu\,I_2(\nu,t)  I_3(\nu,t),   \nonumber \\ [2mm] 
{\cal A}^2 &=&
\ C^{\prime\prime}\int_0^\infty dt\int_0^t d\nu\,I_2(\nu,t) I_4(\nu,t) ,  
  \label{Type_II_amplitude_full}
\end{eqnarray}
where the integrands are given in terms of 
\begin{eqnarray}
I_2(\nu,t)
&:=& (8\pi^2\alpha^\prime t)^{-(p+1)/2}e^{-2\pi r^2 t}\frac{\vartheta_{01}(0,it)^4}{\eta(it)^{12}} \nonumber \\
& & \times \exp\left[ -2\alpha^\prime k^2\left( \pi\frac{\nu(t-\nu)}t+G_B(\nu,t) \right) \right], \\ [2mm] 
I_3(\nu,t) 
&:=& 2\pi^2\frac{(\zeta_1\cdot w)(\zeta_2\cdot w)}{\alpha^\prime}-\zeta_1\cdot\zeta_2\,\partial_\nu^2 G_B(\nu,t), \\ [2mm] 
I_4(\nu,t) 
&:=& 2\pi^2\alpha^\prime\zeta_1\cdot\zeta_2\,k^2G_F(\nu,t)^2. 
\end{eqnarray}
The function $G_B(\nu,t)$ is defined in (\ref{def_G_B}). 
The function $G_F(\nu,t)$ is given in terms of $G_B(\nu,t)$ as 
\begin{equation}
G_F(\nu,t)\ :=\ -\frac1{2\pi}\partial_\nu\Bigl( G_B(\nu,t)-2G_B(\nu/2,t/2) \Bigr). 
   \label{def_G_F}
\end{equation} 
The vector $w^I$ indicates the relative position of the two D-branes. 
For more details of the calculation of the amplitude,  see appendix \ref{app_amplitude}. 

Note that there are two types of transverse scalars distinguished by the relation between $w^I$ and $\zeta^I$. 
We denote by $\phi^\parallel$ the scalar for which $\zeta^I$ is parallel to $w^I$. 
The other scalars for which $\zeta\cdot w=0$ is satisfied are collectively denoted by $\phi^\perp$. 
The following discussions are given for general $\zeta$, and the final results are discussed separately. 

\vspace{5mm}

\subsubsection{The IR part ${\cal A}_{\rm IR}$}

\vspace{5mm}

Let us apply the partial modular transformation to this amplitude with the division point $t=\lambda^{-2}$, as in the previous section. 
We assume $\lambda$ to be small. 

We begin with the analysis of the IR part ${\cal A}_{\rm IR}$. 
As for the bosonic string case, we expect that ${\cal A}_{\rm IR}$ reproduces the amplitude for the low energy effective theory on the brane-antibrane pair. 
Due to the separation of the D-branes, the only low energy field coming from the stretched open string is the tachyon whose mass is $m^2=(r^2-1/2)/\alpha^\prime$. 
This scalar field is circulating in the loop. 

We divide ${\cal A}_{\rm IR}$ into two parts ${\cal A}_{\rm IR}^1$ and ${\cal A}_{\rm IR}^2$, and investigate separately. 
By ignoring exponentially small terms, we obtain for ${\cal A}_{\rm IR}^1$ 
\begin{eqnarray}
{\cal A}_{\rm IR}^1 
&:=& C^{\prime\prime}\int_{\lambda^{-2}}^\infty dt\int_0^t d\nu\,I_2(\nu,t)I_3(\nu,t) \nonumber \\ [2mm] 
&\sim& C^{\prime\prime}\int_{\lambda^{-2}}^\infty dt\int_0^t d\nu\,(8\pi^2\alpha^\prime t)^{-(p+1)/2}e^{-2\pi(r^2-1/2)t}\exp\left( -2\pi\alpha^\prime k^2\frac{\nu(t-\nu)}t \right) \nonumber \\ [2mm] 
& & \times\exp\Bigl( -2\alpha^\prime k^2G_B^0(\nu,t) \Bigr)\left( 2\pi^2\frac{(\zeta_1\cdot w)(\zeta_2\cdot w)}{\alpha^\prime}-\zeta_1\cdot\zeta_2\,\partial_\nu^2G_B^0(\nu,t) \right). 
\end{eqnarray}
In order to compare this expression with field theory amplitudes, we need to extract terms which survive in the zero slope limit. 
This can be achieved by rescaling the integration variables $t$ and $\nu$ as 
\begin{eqnarray}
{\cal A}_{\rm IR}^1 
&\sim& \frac{C^{\prime\prime}}{\alpha^\prime}\int_{\Lambda^{-2}}^\infty d\tau\int_0^\tau d\tau_2\,(8\pi^2\tau)^{-(p+1)/2}e^{-2\pi m^2\tau}\exp\left( -2\pi k^2\frac{\tau_2(\tau-\tau_2)}\tau \right) \nonumber \\ [2mm] 
& & \times\exp\Bigl( -2\alpha^\prime k^2G_B^0(\tau_2/\alpha^\prime,\tau/\alpha^\prime) \Bigr)\left( 2\pi^2\frac{(\zeta_1\cdot w)(\zeta_2\cdot w)}{(\alpha^\prime)^2}-\frac{\zeta_1\cdot\zeta_2}{\alpha^\prime}\,\partial_x^2G_B^0(x,\tau/\alpha^\prime)\Big|_{x=\tau_2/\alpha^\prime} \right). \nonumber \\
   \label{Type_II_IR-1_rescaled}
\end{eqnarray}
By performing a similar calculation done for (\ref{string_contact}) based on the zeta function regularization, we find that ${\cal A}_{\rm IR}^1$ contains a part 
\begin{equation}
\frac{C^{\prime\prime}}{\alpha^\prime}\zeta_1\cdot\zeta_2\int_{\Lambda^{-2}}^\infty d\tau\,(8\pi^2\tau)^{-(p+1)/2}e^{-2\pi m^2\tau}(-2\pi). 
\end{equation}
This can be identified with the field theory amplitude for the Feynman diagram in figure \ref{fig_Feynman_scalar}(b)  calculated in appendix \ref{app_QFT_scalar}, provided 
\begin{equation}
C^{\prime\prime}\ =\ \alpha^\prime g_{\rm YM}^2. 
\end{equation}

When the external states are $\phi^\perp$, the above expression is the only contribution which has a field theory counterpart. 
For $\phi^\parallel$, on the other hand, $\zeta_s^I$ and $w^I$ are given as
\begin{equation}
\zeta_1^I\ =\ \zeta_2^I\ =\ \zeta n^I, \hspace{1cm} w^I\ =\ \frac l\pi n^I
   \label{transverse_scalar}
\end{equation}
for a unit vector $n^I$. 
Since 
\begin{equation}
2\pi^2\frac{(\zeta_1\cdot w)(\zeta_2^I\cdot w)}{(\alpha^\prime)^2}\ =\ 2(2\pi)^2v^2\zeta^2, \hspace{1cm} v\ :=\ \frac l{2\pi\alpha^\prime}, 
\end{equation}
another term in the parenthesis in (\ref{Type_II_IR-1_rescaled}) survives the zero slope limit, and gives
\begin{equation}
g_{\rm YM}^2\int_{\Lambda^{-2}}^\infty d\tau\int_0^\tau d\tau_2\,(8\pi^2\tau)^{-(p+1)/2}e^{-2\pi m^2\tau}\exp\left( -2\pi k^2\frac{\tau_2(\tau-\tau_2)}\tau \right)\cdot 2(2\pi)^2v^2\zeta^2, 
\end{equation}
as long as $v$ is kept finite. 
As explained in appendix \ref{app_QFT_scalar}, this reproduces the amplitude for the Feynman diagram in figure \ref{fig_Feynman_scalar}(a). 

\vspace{5mm}

We observed that ${\cal A}_{\rm IR}^1$ alone reproduces the corresponding field theory amplitudes. 
Then, ${\cal A}_{\rm IR}^2$ should give stringy corrections only. 
Indeed, this can be confirmed by rewriting ${\cal A}_{\rm IR}^2$ as 
\begin{eqnarray}
{\cal A}_{\rm IR}^2 
&:=& \alpha^\prime g_{\rm YM}^2\int_{\lambda^{-2}}^\infty dt\int_0^t d\nu\,I_2(\nu,t)I_4(\nu,t) \nonumber \\ [2mm] 
&\sim& 2\pi^2g_{\rm YM}^2\zeta_1\cdot\zeta_2\,k^2\int_{\Lambda^{-2}}^\infty d\tau\int_0^\tau d\tau_2\,(8\pi^2\tau)^{-(p+1)/2}e^{-2\pi m^2\tau}\exp\left( -2\pi k^2\frac{\tau_2(\tau-\tau_2)}\tau \right) \nonumber \\ [2mm] 
& & \hspace*{2cm}\times \exp\Bigl( -2\alpha^\prime k^2G_B^0(\tau_2/\alpha^\prime,\tau/\alpha^\prime) \Bigr)G_F^0(\tau_2/\alpha^\prime,\tau/\alpha^\prime)^2, 
\end{eqnarray}
where $G_F^0(\nu,t)$ is obtained from $G_F(\nu,t)$ by replacing $G_B(\nu,t)$ in (\ref{def_G_F}) with $G_B^0(\nu,t)$ in (\ref{approx_G_B}). 
Explicitly, it is given as 
\begin{equation}
G_F^0(\nu,t)\ =\ \frac{e^{-\pi\nu}}{1-e^{-2\pi\nu}}-\frac{e^{-\pi(t-\nu)}}{1-e^{-2\pi(t-\nu)}}. 
\end{equation}
Like $G_B^0(\tau_2/\alpha^\prime,\tau/\alpha^\prime)$, the function $G_F^0(\tau_2/\alpha^\prime,\tau/\alpha^\prime)$ localizes the integrand to the regions near $\tau_2=0$ and $\tau_2=\tau$ whose lengths are of order ${\cal O}(\alpha^\prime)$. 
Therefore, each coefficient in the $k^2$-expansion of ${\cal A}_{\rm IR}^2$ is higher order in $\alpha^\prime$, implying that all of them are stringy corrections. 
They are negligible when $\lambda$ is small. 
Thus, in this limit, ${\cal A}_{\rm IR}$  gives the field theory results and the stringy corrections are absorbed into  ${\cal A}_{\rm UV}$ which we will discuss in the following. 

\vspace{5mm}

\subsubsection{The UV part ${\cal A}_{\rm UV}$}
\label{sec:typeIIUV}
\vspace{5mm}

Next, we investigate the UV part ${\cal A}_{\rm UV}$. 
As explained in the previous section, ${\cal A}_{\rm UV}$ contains all the stringy corrections which may be relevant in the low energy effective theory. 
Again, we divide ${\cal A}_{\rm UV}$ into two parts, ${\cal A}_{\rm UV}^1$ and ${\cal A}_{\rm UV}^2$, and investigate them separately. 

\vspace{5mm}

(i) {\it Mass shifts}

\vspace{5mm}

First, we consider 
\begin{equation}
{\cal A}_{\rm UV}^1\ :=\ \alpha^\prime g_{\rm YM}^2\int_{\lambda^2}^\infty ds\int_0^{1/s}d\nu\,\tilde{I}_2(\nu,s)\tilde{I}_3(\nu,s), 
\end{equation}
where 
\begin{eqnarray}
\tilde{I}_2(\nu,s) 
&=& (8\pi^2\alpha^\prime)^{-(p+1)/2}s^{(p-11)/2}e^{-2\pi r^2/s}\frac{\vartheta_{10}(0,is)^4}{\eta(is)^{12}} \nonumber \\
& & \times \exp\left[ -2\alpha^\prime k^2\log\left( -\frac1s\frac{\vartheta_{11}(\nu s,is)}{\eta(is)^3} \right) \right], \\ [2mm] 
\tilde{I}_3(\nu,s) 
&=& 2(2\pi)^2r^2\zeta^2-\zeta^2\left[ 2\pi s+\partial_\nu^2\log\left( -\frac1s\frac{\vartheta_{11}(\nu s,is)}{\eta(is)^3} \right) \right]. 
\label{typeIImassshiftUV-I3}
\end{eqnarray}
This is the expression for $\phi^\parallel$. 
In order to obtain the expression for $\phi^\perp$, we simply ignore the first term in (\ref{typeIImassshiftUV-I3}). 
We expand this expression in terms of $k^2$. 
At order ${\cal O}(k^0)$, we obtain 
\begin{equation}
\alpha^\prime g_{\rm YM}^2(8\pi^2\alpha^\prime)^{-(p+1)/2}\int_{\lambda^2}^\infty ds\,s^{(p-11)/2}e^{-2\pi r^2/s}\frac{\vartheta_{10}(0,is)^4}{\eta(is)^{12}}\Bigl( 2(2\pi)^2r^2\zeta^2s^{-1}-2\pi\zeta^2 \Bigr). 
   \label{typeII_k^0}
\end{equation}
We have performed the $\nu$-integration by using the results in appendix \ref{app_modular}. 

Note that the other part 
\begin{equation}
{\cal A}_{\rm UV}^2\ :=\ \alpha^\prime g_{\rm YM}^2\int_{\lambda^2}^\infty ds\int_0^{1/s} d\nu\,\tilde{I}_2(\nu,s)\tilde{I}_4(\nu,s), 
\end{equation}
where 
\begin{equation}
\tilde{I}_4(\nu,s)\ :=\ 2\pi^2\alpha^\prime\zeta_1\cdot\zeta_2k^2\left( s\sum_{n=1}^\infty \frac{1-\tilde{q}^n}{1+\tilde{q}^n}\sin(2\pi n\nu s) \right)^2, 
\end{equation}
does not have  terms of order ${\cal O}(k^0)$. 

Therefore, the term (\ref{typeII_k^0}) gives the mass shift 
\begin{equation}
\Delta m^2(r)\ =\ -\frac\xi{\alpha^\prime}\int_{\lambda^2}^\infty ds\,s^{(p-11)/2}e^{-2\pi r^2/s}\frac{\vartheta_{10}(0,is)^4}{\eta(is)^{12}}\Bigl( 2(2\pi)^2r^2s^{-1}-2\pi \Bigr) 
   \label{mass_shift_full}
\end{equation}
for $\phi^\parallel$, 
where we defined a dimensionless quantity 
\begin{equation}
\xi\ :=\ (8\pi^2)^{-(p+1)/2}g_{\rm YM}^2(\alpha^\prime)^{(3-p)/2}. 
\end{equation}
Since there is no closed string tachyons, this integral is finite for $p<9$. 
The divergence for $p=9$ is due to a dilaton tadpole. 
The formula (\ref{mass_shift_full}) can be also obtained from the second derivative of the effective potential (\ref{effective_potential}). 

Note that $\Delta m^2(r)$ is actually equal to the physical mass at the 1-loop level since the tree level mass is zero, which implies that the wave function renormalization does not contribute at this level. 

We expand $\Delta m^2(r)$ for large $r$. 
As shown in the previous section, this is obtained by expanding 
\begin{equation}
\frac{\vartheta_{10}(0,is)^4}{\eta(is)^{12}}\ =\ 16+256\,\tilde{q}+{\cal O}(\tilde{q}^2), 
\end{equation}
where $\tilde{q}=e^{-2\pi s}$, and calculate term by term. 
The leading contribution $\Delta m^2_{(0)}(r)$ to the stringy mass shift $\Delta m^2(r)$ for $\phi^\parallel$ is 
\begin{eqnarray}
\Delta m^2_{(0)}(r) 
&=& -16\frac\xi{\alpha^\prime}\int_{\lambda^2}^\infty ds\,s^{(p-11)/2}e^{-2\pi r^2/s}\Bigl( 2(2\pi)^2 r^2s^{-1}-2\pi \Bigr) 
\label{typeIImassshiftleading-para}
 \\ [2mm] 
&\to& -32\pi(8-p)\Gamma({\textstyle \frac{9-p}2})(2\pi r^2)^{(p-9)/2}\frac\xi{\alpha^\prime}, 
\label{typeIImassshiftleading-para2}
\end{eqnarray}
where we have taken the limit $\lambda\to0$ for simplicity. 
We analyze this result for $p<8$ and $p=8$ separately. 
Recall that $r$ is defined as a dimensionless quantity, and related to a dimensionful distance $l$ as $r= l/2\pi\sqrt{\alpha^\prime}.$ 
\begin{itemize}
\item 
This is a negative contribution to the mass shift for $\phi^\parallel$ when $p<8$. 
That is, the transverse scalar, corresponding to the direction of the D-brane separation, becomes tachyonic by the 1-loop stringy corrections. 
Note that the mass shift coming from ${\cal A}_{\rm IR}$ is cancelled in the whole amplitude ${\cal A}$ since the UV cut-off $\Lambda$ introduced by hand should not appear in physical quantities. 
Interestingly, $\Delta m^2_{(0)}(r)$ decreases as $r^{p-9}$ for large $r$. 
On dimensional ground, one might think that the mass shift would decrease as $r^{-2}$. 
Let us compare these two behaviors for $p=3$. 
We obtain 
 \begin{equation}
  \Delta m^2_{(0)}(r) \ \propto\  
  g_{\rm YM}^2 m_{\text str}^{2} (l_{\text str}/l)^6
 \ \ll\ g_{\rm YM}^2 m_{\text str}^{2} (l_{\text str}/l)^2  
 \label{typeIImassp-9}
 \end{equation}
 for $l \gg l_{\rm str}$, 
 where $l_{\rm str}=1/m_{\rm str}=\sqrt{\alpha^\prime}$.
Although the mass shift is not exponentially small, it decreases much faster than the naive expectation when $p<7$. 

\item
It is curious that $\Delta m^2_{(0)}(r)$ vanishes for $p=8$, although this setup might not be interesting for phenomenological purposes. 
For this case, the leading contribution to the mass shift $\Delta m^2_{(1)}(r)$ is exponentially suppressed as a function of $r$ and given by
\begin{eqnarray}
\Delta m^2_{(1)}(r) 
&=& -256\alpha^\prime g_{\rm YM}^2(8\pi^2\alpha^\prime)^{-9/2}\int_{\lambda^2}^\infty ds\,s^{-3/2}e^{-2\pi r^2/s}e^{-2\pi s}\Bigl( 2(2\pi)^2r^2s^{-1}-2\pi \Bigr) \nonumber \\ [2mm] 
&\sim& -2^{10}\sqrt{2}\pi^2e^{-4\pi r}\frac\xi{\alpha^\prime}. 
\end{eqnarray}
See appendix \ref{app_bulk} for details. 
Due to the cancellation of the coefficient of $\Delta m^2_{(0)}(r)$, we obtain a negative mass squared which is {\it exponentially small} compared to the string scale on the D8-$\overline{\rm D8}$ pair. 
\end{itemize}
It is interesting to observe that the absolute value of $\Delta m^2_{(0)}(r)$ is increasing as the D-branes are coming closer to each other. 
For $r^2<\frac12$, the ground state of the stretched string becomes tachyonic, and the D-brane system is unstable for the decay via the tachyon condensation \cite{Sen:2004nf}. 
It seems that the tachyonic mass shift indicates the instability of the system, even after the open string tachyon is regularized by the string tension. 
This might be relevant for phenomenological model building which includes Higgs fields.

There are other transverse scalars $\phi^\perp$. 
The mass shift for them is given by only the second term in (\ref{typeIImassshiftleading-para}) as 
\begin{eqnarray}
\Delta m^2_{(0)}(r) 
&=& -16\frac\xi{\alpha^\prime}\int_{\lambda^2}^\infty ds\,s^{(p-11)/2}e^{-2\pi r^2/s}(-2\pi) \nonumber \\ [2mm] 
&\to& 32\pi\Gamma({\textstyle \frac{9-p}2})(2\pi r^2)^{(p-9)/2}\frac\xi{\alpha^\prime}. 
\end{eqnarray}
This is always positive for $p<8$. 
Thus they do not induce instability of the system. 

\vspace{5mm}

(ii) {\it Wava function renormalizations}

\vspace{5mm}

The wave function renormalization can be also discussed. 
We will only show expressions for $\phi^\parallel$, but the generalization to other scalars $\phi^\perp$ is straightforward. 

We can read off the wave function renormalization from the ${\cal O}(k^2)$ terms. 
From ${\cal A}_{\rm UV}^1$, we obtain 
\begin{eqnarray}
& & 32g_{\rm YM}^2(8\pi^2\alpha^\prime)^{-(p+1)/2}(\alpha^\prime)^2k^2\int_{\lambda^2}^\infty ds\,s^{(p-11)/2}e^{-2\pi r^2/s}\Bigl( (2(2\pi)^2r^2s^{-1}-2\pi)\log s+\pi^2 s \Bigr) \nonumber \\ [2mm] 
&\to& 32\pi^2\Gamma({\textstyle \frac{7-p}2})(2\pi r^2)^{(p-7)/2}\xi k^2+{\cal O}(r^{p-9}),
\end{eqnarray}
where we have taken the limit $\lambda \rightarrow 0$ for simplicity. 
The other part ${\cal A}_{\rm UV}^2$ gives 
\begin{eqnarray}
& & -8\pi^2g_{\rm YM}^2(8\pi^2\alpha^\prime)^{-(p+1)/2}(\alpha^\prime)^2k^2\int_{\lambda^2}^\infty ds\,s^{(p-9)/2}e^{-2\pi r^2/s} \nonumber \\ [2mm] 
&\to& -8\pi^2\Gamma({\textstyle \frac{7-p}2})(2\pi r^2)^{(p-7)/2}\xi k^2. 
\end{eqnarray}
In total, the wave function renormalization is given for large $r$ by 
\begin{equation}
24\pi^2\Gamma({\textstyle \frac{7-p}2})(2\pi r^2)^{(p-7)/2}\xi k^2. 
\end{equation}
In fact, this is divergent for $p\ge7$. 
Similar divergencies appear in the effective potential for D$p$-$\overline{{\rm D}p}$ pair. 
We note that these leading order terms for large $r$ are obtained via the zeta function regularization, while finite terms from the beginning only give sub-leading terms to the wave function renormalization. 
This seems to suggest that they have some stringy origins. 

\vspace{5mm}

It is interesting to perform a corresponding calculation based on Type II supergravity and DBI action. 
This is an extension of the well-known calculation of the effective potential for two D-branes at their ground states. 
For the extension to the wave function renormalization, we calculate the exchange diagrams of the type in figure \ref{fig_closed_channel}. 
In this calculation, we need the interaction vertices between the bulk closed string modes and the scalar fields on the D$p$-brane. 
They are obtained from kinetic terms of the scalar fields in DBI action. 
By using this, we find that the exchange amplitudes cancel among them completely, although the string amplitude gives non-vanishing contributions. 
See appendix \ref{app_DBI} for details. 

Actually, the results from the analysis using DBI action can be understood as follows. 
As explained in appendix \ref{app_DBI}, the Ramond-Ramond field exchange between the D-branes does not contribute to the amplitude for the wave function renormalization. 
Therefore, the result must be the same as the corresponding calculation for parallel D$p$-branes. 
The latter must vanish since the supersymmetric 2-point amplitude vanishes. 
This might suggest that there are some couplings of D-branes to the bulk fields which are missing in DBI action. 
It should be noted, however, that this conclusion is not on a firm footing since the studies in \cite{Pius:2013sca,Pius:2014iaa} do not claim that the stringy calculations of wave function renormalizations, which are not physical observables, are free of ambiguity. 

\vspace{1cm}

\section{Discussion} \label{sec:discuss}

\vspace{5mm}

In this paper, we have investigated various 2-point 1-loop string amplitudes by using the partial modular transformation reviewed in section \ref{sec:PMT}. 
The partial modular transformation can be used as a technique to calculate string amplitudes, similarly to \cite{Iso:2019gbd,Iso:2020ewj} where we calculated the effective potential for a complicated D-brane system. 
We have also observed that the partial modular transformation reveals a structure of the string amplitudes which resembles the amplitudes obtained from the Wilsonian effective action. 
In particular, stringy threshold corrections can be extracted from string amplitudes in intuitively more natural manner. 
For the mass shifts, we reproduced the prescription of \cite{Seiberg:1986ea}. 
We have calculated the mass shifts and the wave function renormalizations for scalar fields on D$p$-$\overline{{\rm D}p}$ pair in Type II string theory. 
The scalar fields are massless at tree level. 
We have found that one of them acquires a tachyonic mass when the distance $r$ of the D-branes are large. 
Interestingly, we have also found that the absolute value of the tachyonic mass is exponentially decreasing with $r$ for $p=8$. 

It will be quite exciting if a hierarchical mass spectrum with an exponentially suppressed stringy threshold corrections could be realized on a D-brane system which is more suitable for phenomenological purposes. 
The model studied in section \ref{sec:typeIIppbar}, however, does not have
such a hierarchy since the exchanges of the massless closed string modes do not cancel in general. 
Therefore, we need to find another D-brane setup where such massless contributions cancel among them. 
As a trial, some modifications of the D$p$-$\overline{{\rm D}p}$ pair are examined in vain in appendix \ref{modified_SS}. 
In the trial we  considered D-branes on a compactified circle whose setup is depicted in figure \ref{fig_SS}
and tried to modify the closed string mass spectrum by inserting non-trivial phases in the sum 
over winding numbers. 
If it would be possible, a hierarchical mass spectrum would be obtained. 
But the trial does not work as it would result in an {\it elimination} of massless closed string modes, especially the graviton. 
Thus there is no hope along this line of approach. 
Instead, we should search for  D-brane setups where massless closed string exchanges {\it cancel} among them
without eliminating the massless closed string states. 
We hope to address other models in future publications. 

\vspace{5mm}

Though the coupling to massless closed string states does not cancel in  the model studied in section \ref{sec:typeIIppbar},
we have an interesting behavior in the coupling. 
The mass shift is given in (\ref{typeIImassp-9}) and proportional to $r^{p-9}$ where $r$ is the distance between D-branes.
This raises a question that what type of interaction between the worldvolume fields on D-branes and the massless closed string states
generates such a mass shift. 
Suppose that one tries to perform calculations for the mass shifts analogous to the ones in appendix \ref{app_DBI}. 
Then, one immediately notice that there is no interaction vertex in DBI action which is available for the calculation. 
Namely, there is no non-derivative coupling involving the scalar fields $\phi$ with massless closed string states such as graviton. 
This is consistent with the above behavior of the mass shift (\ref{mass_shift_full}) obtained from the string amplitude. 
Indeed, if there were a non-derivative coupling of the transverse scalars like $\sqrt{-\det g}\,\phi^2$, 
then we would obtain a mass shift proportional to the graviton propagator, namely the Newtonian potential $r^{p-7}$. 
Since the leading order term $\Delta m^2_{(0)}(r)$ is proportional to $r^{p-9}$, 
we can interpret this result as the absence of the contributions from the ordnary DBI action. 
Then, what kind of modification of the DBI action can possibly reproduce the mass shift $\Delta m^2_{(0)}(r)$? 
One possibility would be the conformal coupling ${\cal R}\phi^2$, where $\cal R$ is the scalar curvature of the target space. 
Since this vertex introduces momentum squared in the exchange amplitude, we can obtain the correct power of $r$. 
There is another reason that the conformal coupling would be suitable. 
Since the transverse scalars are the Nambu-Goldstone bosons, their mass terms are forbidden. 
However, in the presence of other D-branes, they are allowed to acquire masses. 
The influence of the other D-branes are detected through the couplings to the bulk fields. 
Therefore, it is natural that a non-trivial curvature made by the other D-branes induces masses to the scalars. 

It is very important to understand the relation between stringy threshold corrections obtained from string amplitudes and possible modifications of DBI action. 
This would be possible by investigating {\it off-shell} disk amplitudes with insertions of vertex operators for both open string states and closed string states. 
Connecting such disk amplitudes by using propagators of the supergravity fields, we would obtain the large $r$ expansion of a stringy threshold correction. 
The calculations of this kind are expected to be much simpler than the calculations of 1-loop string amplitudes. 
Therefore, this would be a practical tool to search for D-brane systems suitable for phenomenological model building. 

\vspace{5mm}

Finally we would like to address three comments in order. 
The first is a generalization of the partial modular transformation to higher loop amplitudes. 
In this paper, we have observed a similarity between string amplitudes for which the partial modular transformation is performed and amplitudes obtained from Wilsonian effective action. 
It is quite interesting to check whether this similarity persists in higher loop amplitudes. 
Since there are several moduli, we should perform the partial modular transformation separately, and the resulting structure would be much more complicated. 
However, the advantage of this way of understanding is obvious. 
Assuming the similarity persists for all loops, which seems intuitively valid, the partial modular transformation gives us a method to obtain the Wilsonian effective action of string theory, which enables us to perform any off-shell calculations at least all orders of string perturbation theory. 

The second comment is about the regularization of string amplitudes. 
In this paper, we have used the zeta function regularization for actual calculations on string amplitudes. 
The validity of this regularization scheme can be justified by the observations that the obtained results are reasonable. 
However, it is desired to understand the regularization of string amplitudes more systematically. 
Recently, there appeared \cite{Eberhardt:2023xck} whose aim is a realization of the proposal in \cite{Witten:2013pra}. 
The paper also analyzes mass shifts. 
It is very interesting to find a possible relation between \cite{Eberhardt:2023xck} and the present work. 
Since our viewpoint has a direct connection to ordinary quantum field theories, it would be reasonable to expect that the understanding of the regularization issues would become comparably easier. 

The third comment is a generalization to higher-point amplitudes. 
Needless to say, the application of the partial modular transformation is not limited to 2-point amplitudes. 
By generalizing to higher point amplitudes, we can investigate stringy threshold corrections to coupling constants, especially at and above the string scale. 
The corrections to the gauge couplings can even be extracted from the 2-point amplitude via the wave function renormalization. 
These issues will be hopefully reported elsewhere. 

\vspace{1cm}

{\bf Acknowledgements}

\vspace{5mm}

We would like to thank Yuta Hamada and Sota Nakajima for valuable comments. 
This work is supported in part by Grants-in-Aid for Scientific Research 
No.18H03708 and No.19K03851 from the Japan Society for the Promotion of Science.
NK would like to thank the members of KEK theory center for their hospitality where we could gather and make fruitful discussions in person
under COVID-19.

\appendix

\vspace{1cm}

\section{String 1-loop amplitudes}
\label{app_amplitude}

\vspace{5mm}

In this appendix, we summarize various string 1-loop ampitudes discussed in this paper. 
We basically follow \cite{Green:1987sp,Green:1987mn}. 

\vspace{5mm}

\subsection{Bosonic string amplitudes}

\vspace{5mm}

We begin with bosonic open string amplitudes. 
The open string is attached to a pair of parallel D$p$-branes. 
We assume that $X^\mu$ with $\mu=0,\cdots,p-1$ correspond to the parallel directions of the D$p$-branes, and $X^I$ with $I=p,\cdots,25$ correspond to the transverse directions which satisfy 
\begin{equation}
X^I(\tau,\sigma+\pi)\ =\ X^I(\tau,\sigma)+\pi w^I, 
\end{equation}
where the vector $w^I$ describes the relative position of the D$p$-branes. 
More precisely, if the distance between the D$p$-branes is $l$ and the relative direction is given by a unit vector $n^I$, then $w^I$ is given as 
\begin{equation}
w^I\ =\ \frac l\pi n^I. 
\end{equation}

The schematic form of 2-point 1-loop amplitudes is 
\begin{equation}
{\cal A}\ =\ (2\pi)^2\int_0^\infty dt_1\int_0^\infty dt_2\,{\rm Tr}_X\Bigl( e^{-2\pi t_1(L_0^X-1)}V_1\,e^{-2\pi t_2(L_0^X-1)}V_2 \Bigr), 
   \label{app_amplitude_schematic}
\end{equation}
where $L_0^X$ is given in terms of the zero modes $\hat{p}^\mu$ and the non-zero modes $\hat{\alpha}_n^\mu$ and $\hat{\alpha}_n^I$ of $X^\mu(\tau,\sigma)$ and $X^I(\tau,\sigma)$ by 
\begin{equation}
L_0^X\ =\ \frac{l^2}{4 \pi^2 \alpha^\prime} + \alpha^\prime \hat{p}^2 + \sum_{n >0} \hat\alpha_{-n}\cdot \hat\alpha_n. 
\end{equation}
The external states are specified by the vertex operators $V_s$ with $s=1,2$. 
To simplify calculations, we follow \cite{Green:1987sp} and use 
\begin{equation}
V_s\ :=\ c_t\exp\Bigl( ik_{s,\mu} X^\mu+ic_\epsilon\epsilon_{s,\mu}\partial_\tau X^\mu+c_\zeta\zeta_{s,I}\partial_\sigma X^I \Bigr)\Big|_{\tau=\sigma=0}. 
   \label{app_vertex_bosonic}
\end{equation}
The normal ordering is assumed, as usual. 
Here, $\epsilon_s$ are polarization vectors for the external gauge bosons, $\zeta_s$ are vectors specifying the transverse scalars, and $c_t$, $c_\epsilon$ and $c_\zeta$ are some constants. 
We impose the transversality conditions $k_s\cdot\epsilon_s=0$. 
Note that the factor $i$ in front of $\epsilon_s$ is due to the Wick rotation of $\tau$. 

We can obtain the amplitude with the desired external states from (\ref{app_amplitude_schematic}) and (\ref{app_vertex_bosonic}) as follows. 
In order to obtain the tachyon amplitude, we simply set $\epsilon_s=0$ and $\zeta_s=0$. 
The gauge boson amplitude can be obtained by expanding ${\cal A}$ in terms of $\epsilon_s$ and picking up terms proportional to $\epsilon_{1,\mu}\epsilon_{2,\nu}$. 
Likewise, the coefficients of $\zeta_{1,I}\zeta_{2,J}$ give the amplitude for the transverse scalars. 

Note that we included numerical factors $c_t, c_\epsilon$ and $c_\zeta$ in the above expression of $V_s$ in order to match with the corresponding field theory amplitudes. 
Typically, they can be determined by examining tree level amplitudes and comparing them with field theory amplitudes. 
In this paper, since all of the factors appear as overall coefficients of the string 1-loop amplitudes, we simply adjust the overall coefficients by comparing them with the corresponding field theory 1-loop amplitudes. 

The calculation of the trace in (\ref{app_amplitude_schematic}) can be performed separately for the zero modes and the non-zero modes of $X^\mu$ and $X^I$. 
The zero modes $\hat{x}^\mu$ and $\hat{p}^\mu$ of $X^\mu(\tau,\sigma)$ and $w^I$ in $X^I(\tau,\sigma)$ give 
\begin{eqnarray}
& & (2\pi)^{26}\delta^{26}(k_1+k_2)(8\pi^2\alpha't)^{-(p+1)/2}e^{-2\pi r^2t}\exp\left( -2\pi\alpha'k^2\frac{\nu(t-\nu)}t \right) \nonumber \\
& & \times \exp\left( \frac{\alpha'}{\pi t}c_\epsilon^2\epsilon_1\cdot\epsilon_2
+c_\zeta\zeta_1\cdot w+c_\zeta\zeta_2\cdot w \right), 
   \label{app_zeromode}
\end{eqnarray}
where 
\begin{equation}
t\ :=\ t_1+t_2, \hspace{1cm} \nu\ :=\ t_2, 
\end{equation}
and we have used the momentum conservation $k_1=-k_2=k$. 
The quantity $r^2$ is obtained from $w^I$ as 
\begin{equation}
r^2\ =\ \frac{w^2}{4\alpha'}\ =\ \frac{l^2}{4\pi^2\alpha'}. 
\end{equation}
In the above expression of (\ref{app_zeromode}), 
we have ignored terms proportional to $\epsilon_s^2$ since they are irrelevant in obtaining the amplitudes
(e.g., 2-point gauge field amplitude is proportional to $\epsilon_1\cdot\epsilon_2$.)

A contribution from each non-zero mode $\hat{\alpha}_n^\mu$ and $\hat{\alpha}_n^I$ can be calculated straightforwardly. 
Including the ghost contributions, we obtain 
\begin{equation}
\eta(it)^{-24}\exp\Bigl( -2\alpha'k^2G_B(\nu,t) \Bigr)\exp\left( -\frac{2\alpha'}{(2\pi)^2}(c_\epsilon^2\epsilon_1
\cdot\epsilon_2+c_\zeta^2\zeta_1\cdot\zeta_2)\partial_\nu^2G_B(\nu,t) \right), 
   \label{app_nonzeromode}
\end{equation}
where again we only include terms relevant to the amplitudes. 
The function $G_B(\nu,t)$ is defined as 
\begin{equation}
G_B(\nu,t)\ :=\ \log\left[ (1-e^{-2\pi\nu})\prod_{m=1}^\infty\frac{(1-e^{-2\pi\nu}q^m)(1-e^{2\pi\nu}q^m)}{(1-q^m)^2} \right], \hspace{1cm} q\ :=\ e^{-2\pi t}. 
\end{equation}
The product of (\ref{app_zeromode}) and (\ref{app_nonzeromode}) give the integrand of (\ref{app_amplitude_schematic}). 

To summarize, the explicit forms of the amplitudes, up to overall coefficients, are given as 
\begin{eqnarray}
{\cal A} 
&\propto& (2\pi)^2\int_0^\infty dt\int_0^t d\nu\,(8\pi^2\alpha't)^{-(p+1)/2}e^{-2\pi r^2t}\eta(it)^{-24}\exp\left[ -2\alpha'k^2\left( \pi\frac{\nu(t-\nu)}t+G_B(\nu,t) \right) \right] \nonumber \\ [2mm] 
& & \hspace*{1cm}\times \left\{
\begin{array}{lc}
1, & ({\rm tachyon}) \\ [3mm] 
{\displaystyle 2\alpha'\epsilon_1\cdot\epsilon_2\left( \frac1{2\pi t}-\frac1{(2\pi)^2}\partial_\nu^2G_B(\nu,t) \right),} & (\mbox{gauge boson}) \\ [4mm] 
{\displaystyle (\zeta_1\cdot w)(\zeta_2\cdot w)-\frac{2\alpha'}{(2\pi)^2}\zeta_1\cdot\zeta_2\partial_\nu^2G_B(\nu,t)}, & (\mbox{scalar}) 
\end{array}
\right.
   \label{app_amplitude_explicit}
\end{eqnarray}
where the factor $(2\pi)^{26}\delta^{26}(k_1+k_2)$ has been removed. 
The overall coefficients will be fixed by comparing with the corresponding field theory amplitudes. 

\vspace{5mm}

\subsection{Type II superstring amplitudes}

\vspace{5mm}

In Type II string theory, we mainly consider a 1-loop amplitude for the open string stretched between a D$p$-brane and a $\overline{{\rm D}p}$-brane, whose external states are transverse scalars. 
In the Green-Schwarz formalism, the schematic form of the amplitude is 
\begin{equation}
{\cal A}\ =\ (2\pi)^2\int_0^\infty dt_1\int_0^\infty dt_2\,{\rm STr}\Bigl( e^{-2\pi t_1(L_0-1/2)}V_1\,e^{-2\pi t_2(L_0-1/2)}V_2\ \Bigr), 
\end{equation}
where $L_0$ is given as 
\begin{equation}
L_0\ :=\ L_0^X+L_0^S, \hspace{1cm} L_0^S\ :=\ \sum_{r\ge\frac12}r\hat{S}_{-r}^a\hat{S}^a_r. 
\end{equation}
We have used the fact that, for the D$p$-$\overline{{\rm D}p}$ pair, it turns out that $S^a$ satisfy the NS boundary condition, as explained in appendix \ref{app_GS}. 

The vertex operators for transverse scalars are given as 
\begin{equation}
V_s\ :=\ c\,\zeta_{s,I}\left( \partial_\sigma X^I+\frac{\alpha'}2S^a\gamma^{I\mu}_{ab}S^bk_{s,\mu} \right)e^{ik_{s,\mu} X^\mu}\Big|_{\tau=\sigma=0}, 
\end{equation}
where the light cone components $k_\pm$ are set to zero. 
The change in the relative sign compared to the one in \cite{Green:1987sp} comes from the T-duality as follows. 
As explained in subsection \ref{sec:typeII_GS}, the vertex operators $V_s$ above are obtained from (\ref{vertex_typeII_Neumann}) by taking T-dualities along the transverse directions. 
For the fermionic fields $S^a$, this amounts to 
\begin{equation}
S^a\ \to\ (\gamma^\perp S)^a, \hspace{1cm} \gamma^\perp\ :=\ \prod_I\gamma^I.  
\end{equation}
See appendix \ref{boundary_condition_GS}. 
Then, the fermion bilinear term in $V_s$ becomes 
\begin{equation}
S\gamma^{I\mu}S\ \to\ S(\gamma^\perp)^T\gamma^{I\mu}\gamma^\perp S\ =\ -S\gamma^{I\mu}S. 
\end{equation}

The supertrace can be written as 
\begin{eqnarray}
& & q^{-1/2}\,{\rm Tr}_X\Bigl(e^{-2\pi t_1L_0^X}(\zeta_1\partial_\sigma X)\,e^{ik_1X}e^{-2\pi t_1L_0^X}(\zeta_2\partial_\sigma X) e^{ik_2X} \Bigr){\rm STr}_S\Bigl( e^{-2\pi tL_0^S} \Bigr) \nonumber \\ [2mm] 
&+& \frac14(\alpha')^2q^{-1/2}\,{\rm Tr}_X\Bigl(e^{-2\pi t_1L_0^X}e^{ik_1X}e^{-2\pi t_1L_0^X}e^{ik_2 X} \Bigr) \nonumber \\ [2mm] 
& & \times{\rm STr}_S\Bigl( e^{-2\pi t_1L_0^S}(\zeta_1S\gamma Sk_1) e^{ik_1 X}e^{-2\pi t_2L_0^S}(\zeta_2S\gamma Sk_2) e^{ik_2 X} \Bigr),
\end{eqnarray}
where we set $c=1$ for simplicity. 
This shows that ${\cal A}$ can be obtained from the tachyon amplitude and the scalar amplitude in  (\ref{app_amplitude_explicit}) by multiplying traces for fermions $S^a$. 
The necessary traces are given as 
\begin{equation}
{\rm Str}_S\Bigl( e^{-2\pi tL_0^S} \Bigr)\ =\ \prod_{m=1}^\infty(1-q^{m-1/2})^8 
\end{equation}
and 
\begin{eqnarray}
& & {\rm Str}_S\Bigl( e^{-2\pi t_1L_0^S}(\zeta_1S\gamma Sk_1) e^{ik_1\cdot X}e^{-2\pi t_2L_0^S}(\zeta_2S\gamma Sk_2) e^{ik_2\cdot X} \Bigr) \nonumber \\
&=& 4\zeta_1\cdot\zeta_2 k^2G_F(\nu,t)^2\prod_{m=1}^\infty(1-q^{m-1/2})^8, 
\end{eqnarray}
where the function $G_F(\nu,t)$ is defined as 
\begin{equation}
G_F(\nu,t)\ :=\ -\frac1{2\pi}\partial_\nu\Bigl( G_B(\nu,t)-2G_B(\nu/2,t/2) \Bigr). 
\label{GF-by-GB}
\end{equation}
This is the propagator for an NS fermion on the annulus. 
This can be seen by rewriting $G_F(\nu,t)$ as 
\begin{equation}
G_F(\nu,t)\ =\ \sum_{m\in\mathbb{Z}}\frac{e^{-\pi\nu}q^{m/2}}{1-e^{-2\pi\nu}q^m}. 
\end{equation}
Due to the summation, $G_F(\nu,t)$ is periodic in $\nu$ with period $t$. 

Now, we obtain the amplitude as 
\begin{equation}
{\cal A}\ \propto\ {\cal A}_1+{\cal A}_2
\end{equation}
where  
\begin{eqnarray}
{\cal A}_1 
&=& (2\pi)^2\int_0^\infty \int_0^t d\nu\,(8\pi^2\alpha't)^{-(p+1)/2}e^{-2\pi r^2t}\frac{\vartheta_{01}(0,it)^4}{\eta(it)^{12}}\exp\left[ -2\alpha'k^2\left( \pi\frac{\nu(t-\nu)}t+G_B(\nu,t) \right) \right] \nonumber \\ [2mm] 
& & \hspace*{2cm}\times\left( (\zeta_1\cdot w)(\zeta_2\cdot w)-\frac{2\alpha'}{(2\pi)^2}\partial_\nu^2G_B(\nu,t) \right), \\ [2mm] 
{\cal A}_2 
&=& (2\pi)^2\int_0^\infty \int_0^t d\nu\,(8\pi^2\alpha't)^{-(p+1)/2}e^{-2\pi r^2t}\frac{\vartheta_{01}(0,it)^4}{\eta(it)^{12}}\exp\left[ -2\alpha'k^2\left( \pi\frac{\nu(t-\nu)}t+G_B(\nu,t) \right) \right] \nonumber \\ [2mm] 
& & \hspace*{2cm}\times (\alpha')^2\zeta_1\cdot\zeta_2 k^2G_F(\nu,t)^2.  
\end{eqnarray}

\vspace{1cm}

\section{Modular transformation} \label{app_modular}

\vspace{5mm}

In this paper, we use various functions. 
To fix conventions, we show their definitions in the following.   
\begin{eqnarray}
\eta(\tau) 
&:=& q^{1/24}\prod_{m=1}^\infty(1-q^m), \\
\vartheta_{00}(v,\tau) 
&:=& \prod_{m=1}^\infty(1-q^m)(1+zq^{m-1/2})(1+z^{-1}q^{m-1/2}), \\
\vartheta_{01}(v,\tau) 
&:=& \prod_{m=1}^\infty(1-q^m)(1-zq^{m-1/2})(1-z^{-1}q^{m-1/2}), \\
\vartheta_{10}(v,\tau) 
&:=& 2q^{1/8}\cos\pi v\prod_{m=1}^\infty(1-q^m)(1+zq^m)(1+z^{-1}q^m), \\
\vartheta_{11}(v,\tau) 
&:=& -2q^{1/8}\sin\pi v\prod_{m=1}^\infty(1-q^m)(1-zq^m)(1-z^{-1}q^m), 
\end{eqnarray}
where 
\begin{equation}
q\ :=\ e^{2\pi i\tau}, \hspace{1cm} z\ :=\ e^{2\pi iv}. 
\end{equation}
Various formulas for them can be found in \cite{Polchinski:1998rq}. 

\vspace{5mm}

Recall the definition of $G_B(\nu,t)$ given as 
\begin{equation}
G_B(\nu,t)\ :=\ \log\left[ (1-e^{-2\pi\nu})\prod_{m=1}^\infty\frac{(1-e^{-2\pi\nu}q^m)(1-e^{2\pi\nu} q^m)}{(1-q^m)^2} \right], \hspace{1cm} \tau\ =\ it. 
\end{equation}
We find that the right-hand side can be rewritten as 
\begin{equation}
\log\left[ (1-e^{-2\pi\nu})\prod_{m=1}^\infty\frac{(1-e^{-2\pi\nu}q^m)(1-e^{2\pi\nu} q^m)}{(1-q^m)^2} \right]
 =\ \log\left( ie^{-\pi\nu}\frac{\vartheta_{11}(i\nu,it)}{\eta(it)^3} \right). 
\end{equation}
The modular transformation of $G_B(\nu,t)$ is therefore determined by the following transformation formulas  \begin{equation}
\vartheta_{11}(i\nu,it)\ =\ i s^{1/2} e^{\pi \nu^2/t} \vartheta_{11}(\nu s, is), \hspace{1cm}  
\eta(it) = s^{1/2} \eta(is),  
\end{equation}
found in e.g. \cite{Polchinski:1998rq}, where $s=1/t$. 
We find 
\begin{equation}
\log\left( ie^{-\pi\nu}\frac{\vartheta_{11}(i\nu,it)}{\eta(it)^3} \right)\ =\ \log\left( -\frac1s\frac{\vartheta_{11}(\nu s,is)}{\eta(is)^3} \right)-\pi\nu+\pi\nu^2 s. 
\end{equation}
This implies 
\begin{equation}
\pi\frac{\nu(t-\nu)}t+G_B(\nu,t)\ =\ \log\left( -\frac1s\frac{\vartheta_{11}(\nu s,is)}{\eta(is)^3} \right). 
\label{log-theta11}
\end{equation}

\vspace{5mm}

It is useful to obtain the Fourier expansion of the right-hand side of (\ref{log-theta11}). 
This is obtained as follows. 
We find 
\begin{eqnarray}
\log\left( -\frac1s\frac{\vartheta_{11}(\nu s,is)}{\eta(is)^3} \right) 
&=& \sum_{m=1}^\infty \log(1-e^{2\pi i\nu s}\tilde{q}^m)(1-e^{-2\pi i\nu s}\tilde{q}^m) \nonumber \\
& & +\log(2\sin\pi\nu s)-\log\left( s\prod_{m=1}^\infty(1-\tilde{q}^m)^2 \right), 
\end{eqnarray}
where $\tilde{q} := e^{-2 \pi s}$. 
The first sum can be rewritten as 
\begin{eqnarray}
\sum_{m=1}^\infty \log(1-e^{2\pi i\nu s}\tilde{q}^m)(1-e^{-2\pi i\nu s}\tilde{q}^m) 
&=& -2\sum_{m=1}^\infty\sum_{n=1}^\infty\frac1n\tilde{q}^{nm}\cos(2\pi n\nu s) \nonumber \\ [2mm] 
&=& -\sum_{n=1}^\infty\frac1n\frac{2\tilde{q}^n}{1-\tilde{q}^n}\cos(2\pi n\nu s). 
\end{eqnarray}
The second term can be similarly rewritten as 
\begin{eqnarray}
\log(2\sin\pi\nu s) 
&=& \frac12\log(1-e^{2\pi i\nu s})(1-e^{-2\pi i\nu s}) \nonumber \\ [2mm] 
&=& -\sum_{n=1}^\infty\frac1n\cos(2\pi n\nu s). 
\end{eqnarray}
Then we obtain the desired formula 
\begin{equation}
\log\left( -\frac1s\frac{\vartheta_{11}(\nu s,is)}{\eta(is)^3} \right)\ =\ -\sum_{n=1}^\infty\frac1n\frac{1+\tilde{q}^n}{1-\tilde{q}^n}\cos(2\pi n\nu s)-\log\left( s\prod_{m=1}^\infty(1-\tilde{q}^m)^2 \right). 
   \label{app_G_F_Fourier}
\end{equation}
The Fourier expansion of $G_F(\nu,t)$ defined in (\ref{def_G_F})
 is given by
\begin{equation}
G_F(\nu,t)\ =\ s\sum_{n=1}^\infty\frac{1-\tilde{q}^n}{1+\tilde{q}^n}\sin(2\pi n\nu s) 
\end{equation}
which can be derived from (\ref{app_G_F_Fourier}), (\ref{log-theta11}) and (\ref{GF-by-GB}). 

\vspace{1cm}

\section{Field theory amplitudes} \label{app_QFT}

\vspace{5mm}

In this appendix, we summarize various 1-loop amplitudes in quantum field theories which are discussed in this paper. 

\vspace{5mm}

\subsection{2-point amplitude for gauge bosons} \label{app_QFT_gauge}

\vspace{5mm}
 
\begin{figure}
\begin{center}
\includegraphics{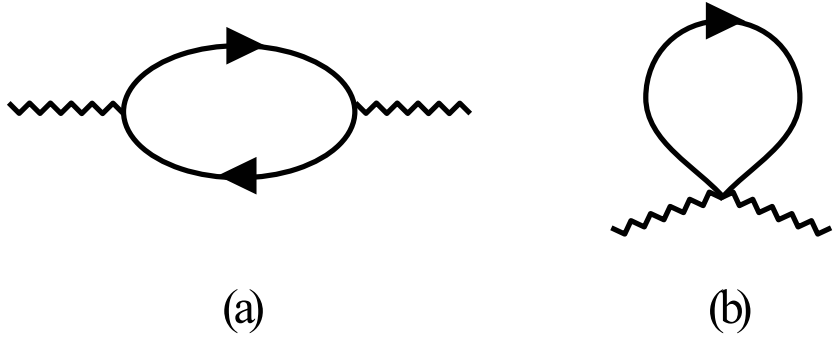}
\end{center}
\caption{Feynman diagrams for 2-point 1-loop amplitudes of gauge bosons in scalar QED. 
The wavy lines represent gauge bosons and the solid lines with arrows represent charged scalar fields. }
   \label{fig_Feynman_gauge}
\end{figure}

For the Feynman diagram in figure \ref{fig_Feynman_gauge}(a), the 2-point amplitude of gauge fields in scalar  QED is given by 
\begin{eqnarray}
& & \frac{g_{\rm YM}^2}{2i}\int \frac{d^Dp}{(2\pi)^D}\frac{4(\epsilon_1\cdot p)(\epsilon_2\cdot p)}{(p^2+m^2)((p-k)^2+m^2)} \nonumber \\ [2mm] 
&=& \frac{g_{\rm YM}^2}{2i}\epsilon_1^\mu\epsilon_2^\nu(2\pi)^2\int_0^\infty d\tau_1\int_0^\infty d\tau_2\int\frac{d^Dp}{(2\pi)^D}\, p_\mu p_\nu \,\exp\Bigl( -2\pi \tau_1(p^2+m^2)-2\pi \tau_2((p-k)^2+m^2) \Bigr) \nonumber \\ [2mm] 
&=& g_{\rm YM}^2\epsilon_1\cdot\epsilon_2\int_0^\infty d\tau\int_0^\tau d\tau_2\,(8\pi^2\tau)^{-D/2}e^{-2\pi m^2\tau}\exp\left( -2\pi k^2\frac{\tau_2(\tau-\tau_2)}\tau \right)\cdot \frac{2\pi}\tau, 
\label{C1gaugeamplitude}
\end{eqnarray}
where $\tau:=\tau_1+\tau_2$. 
We introduce a UV cut-off for the $\tau$-integral by the following replacement: 
\begin{equation}
\int_0^\infty d\tau \hspace{5mm} \Rightarrow \hspace{5mm} \int_{\Lambda^{-2}}^\infty d\tau. 
\end{equation}

The other Feynman diagram in figure \ref{fig_Feynman_gauge}(b) corresponds to the contact interaction and the amplitude is 
\begin{eqnarray}
& & -\frac{g_{\rm YM}^2}{i}\int\frac{d^Dp}{(2\pi)^D}\frac{\epsilon_1\cdot\epsilon_2}{p^2+m^2} \nonumber \\ [2mm] 
&=& ig_{\rm YM}^2\epsilon_1\cdot\epsilon_2 (2\pi)\int_{\Lambda^{-2}}^\infty d\tau\int\frac{d^Dp}{(2\pi)^D}\,\exp\Bigl( -2\pi \tau(p^2+m^2) \Bigr) \nonumber \\ [2mm] 
&=& g_{\rm YM}^2\epsilon_1\cdot\epsilon_2\int_{\Lambda^{-2}}^\infty d\tau\,(8\pi^2\tau)^{-D/2}e^{-2\pi m^2\tau}\cdot(-2\pi). 
   \label{app_amplitude_contact}
\end{eqnarray}

In the above expressions, the Schwinger parameters are dimensionful. 
In order to compare with string amplitudes, we define dimensionless parameters as 
\begin{equation}
t\ :=\ \frac\tau{\alpha'}, \hspace{1cm} \nu\ :=\ \frac{\tau_2}{\alpha'}. 
\end{equation}
In terms of $t$ and $\nu$, the amplitudes become 
\begin{eqnarray}
& & \alpha'g_{\rm YM}^2\epsilon_1\cdot\epsilon_2\int_{\Lambda^{-2}/\alpha'}^\infty dt\int_0^t d\nu\,(8\pi^2\alpha't)^{-D/2}e^{-2\pi m^2\alpha't}\exp\left( -2\pi \alpha'k^2\frac{\nu(t-\nu)}t \right)\cdot \frac{2\pi}t, \nonumber \\ 
   \label{app_QFT_gauge1} \\
& & \alpha'g_{\rm YM}^2\epsilon_1\cdot\epsilon_2\int_{\Lambda^{-2}/\alpha'}^\infty dt\,(8\pi^2\alpha't)^{-D/2}e^{-2\pi m^2\alpha't}\cdot(-2\pi). 
\end{eqnarray}

\vspace{5mm}

\subsection{2-point amplitude for tachyons} \label{app_QFT_tachyon}

\vspace{5mm}

\begin{figure}
\begin{center}
\includegraphics{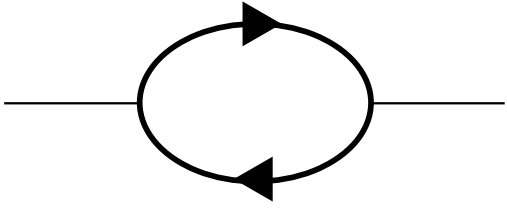}
\end{center}
\caption{The Feynman diagram for 2-point amplitude of tachyons on a D$p$-brane. 
The solid lines with arrows represent the charged tachyon, and the thin solid lines represent the neutral tachyon on a D$p$-brane. }
   \label{fig_Feynman_tachyon}
\end{figure}

The part of the action for the tachyons which is relevant for the discussion in section \ref{sec:renorm} is given by 
\begin{equation}
S\ =\ \int d^D x\, \Bigl( -|\partial_\mu T |^2 + g T_0 |T|^2+\cdots \Bigr) 
\end{equation}
where $T_0$ is a real tachyon field coming from an open string living on one D-brane, and $T$ is a complex 
tachyon field coming from the stretched open string between D-branes.   
For the Feynman diagram in figure \ref{fig_Feynman_tachyon}, the amplitude is 
\begin{eqnarray}
& & \frac{g^2}i\int\frac{d^Dp}{(2\pi)^D}\frac1{p^2+m^2}\frac1{(p-k)^2+m^2} \nonumber \\ [2mm] 
&=& (2\pi)^2g^2\int_{\Lambda^{-2}}^\infty d\tau\int_0^\tau d\tau_2\,(8\pi^2\tau)^{-D/2}e^{-2\pi m^2\tau}\exp\left( -2\pi k^2\frac{\tau_2(\tau-\tau_2)}\tau \right). 
    \label{app_amplitude_tachyon1}
\end{eqnarray}
In the rescaled variables, this becomes 
\begin{equation}
(2\pi)^2g^2(\alpha')^2\int_{\Lambda^{-2}/\alpha'}^\infty dt\int_0^t d\nu\,(8\pi^2\alpha't)^{-D/2}e^{-2\pi m^2\alpha't}\exp\left( -2\pi \alpha'k^2\frac{\nu(t-\nu)}t \right). 
   \label{app_amplitude_tachyon}
\end{equation}

\vspace{5mm}

\subsection{2-point amplitude for transverse scalars} \label{app_QFT_scalar}

\vspace{5mm}

\begin{figure}
\begin{center}
\includegraphics{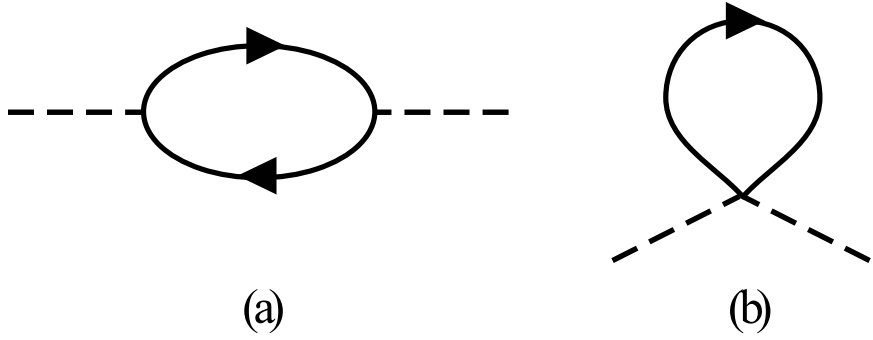}
\end{center}
\caption{Feynman diagrams for 2-point amplitudes of transverse scalars in the low energy effective action of a brane-antibrane pair. 
The dashed lines represent the transverse scalar, and the solid lines with arrows represent the charged scalar. 
Note that the coupling constant for (a) is given by $v$ defined in (\ref{scalar_vev}). }
   \label{fig_Feynman_scalar}
\end{figure}

First, we need to specify the worldvolume effective action for D$p$-$\overline{{\rm D}p}$ pair. 
For the case $p=9$, the relevant part of the action is 
\begin{equation}
S_9\ =\ \frac1{g_{\rm YM}^2}\int d^{10}x\left[ -\frac14F_{MN}^2-\frac14\bar{F}_{MN}^2-|\partial_MT-i(A_M-\bar{A}_M)T|^2-\frac1{2\alpha'}|T|^2+\cdots \right], 
\end{equation}
where $A_M$ ($\bar{A}_M$) is the gauge field on the D9-brane ($\overline{\rm D9}$-brane), respectively, and $T$ is a tachyon coming from the open string stretched between the D-branes. 
The D-brane system for a lower $p$ can be obtained via the T-duality. 
At the level of the effective action, it amounts to performing the dimensional reduction \cite{Taylor:1997dy}. 
The part of the effective action which is necessary to obtain a 2-point amplitude for a transverse scalar $\phi$ on the D$p$-brane is 
\begin{equation}
S_p\ =\ \frac1{g_{\rm YM}^2}\int d^{p+1}x\left[ -\frac12(\partial_\mu\phi)^2-\frac12(\partial_\mu\bar{\phi})^2-(\phi-\bar{\phi})^2|T|^2-\frac1{2\alpha'}|T|^2+\cdots \right], 
\end{equation}
where $\bar{\phi}$ is another transverse scalar field on the $\overline{\rm D9}$-brane corresponding to the same direction as $\phi$. 

Suppose that $\phi$ and $\bar{\phi}$ have vevs $\phi_0$ and $\bar{\phi}_0$, respectively. 
We have 
\begin{equation}
\phi\ =\ \phi_0+\varphi, \hspace{1cm} \bar{\phi}\ =\ \bar{\phi}_0+\bar{\varphi}. 
\end{equation}
Then, the interaction vertices for the fluctuations $\varphi,\bar{\varphi}$ are 
\begin{equation}
(\phi-\bar{\phi})^2|T|^2\ =\ v^2|T|^2+2v(\varphi-\bar{\varphi})|T|^2+(\varphi-\bar{\varphi})^2|T|^2, 
\end{equation}
where $v:=\phi_0-\bar{\phi}_0$. 
In order to reproduce the correct tachyon mass $m^2=(r^2-\frac12)/\alpha'$, we should take 
\begin{equation}
v^2\ =\ \frac{r^2}{\alpha'}\ =\ \left( \frac l{2\pi\alpha'} \right)^2. 
   \label{scalar_vev}
\end{equation}

Now, we calculate 2-point amplitudes. 
The Feynman diagram in figure \ref{fig_Feynman_scalar}(b) gives 
\begin{equation}
-\frac{g_{\rm YM}^2}{i}\int\frac{d^Dp}{(2\pi)^D}\frac1{p^2+m^2}\ =\ g_{\rm YM}^2\int_{\Lambda^{-2}}^\infty d\tau\,(8\pi^2\tau)^{-D/2}e^{-2\pi m^2\tau}(-2\pi). 
\end{equation}
The other Feynman diagram in figure \ref{fig_Feynman_scalar}(a) gives 
\begin{eqnarray}
& & \frac{g_{\rm YM}^2}{2i}(2v)^2\int\frac{d^Dp}{(2\pi)^D}\frac1{p^2+m^2}\frac1{(p-k)^2+m^2} \nonumber \\ [2mm] 
&=& g_{\rm YM}^2\int_{\Lambda^{-2}}^\infty d\tau\int_0^\tau d\tau_2\,(8\pi^2\tau)^{-D/2}e^{-2\pi m^2\tau}\exp\left( -2\pi k^2\frac{\tau_2(\tau-\tau_2)}\tau \right)\cdot2(2\pi)^2v^2. \nonumber \\
\end{eqnarray}

\vspace{1cm}

\section{Leading order terms in string amplitudes} \label{app_leading}

\vspace{5mm}

In this appendix, we evaluate the integral 
\begin{equation}
\int_0^t d\nu\,\exp\left[ -2\alpha'k^2\left( \pi\frac{\nu(t-\nu)}t+G_B(\nu,t) \right) \right]\left( \frac{2\pi}t-\partial_\nu^2G_B(\nu,t) \right), 
\end{equation}
which appears in the 2-point amplitude for the gauge bosons discussed in section \ref{sec:PMT}. 
As explained there, we can make the following replacement 
\begin{equation}
G_B(\nu,t)\ \to\ G_B^0(\nu,t)\ :=\ \log(1-e^{-2\pi\nu})+\log(1-e^{-2\pi(t-\nu)}) 
\end{equation}
in the leading order approximation. 

We expand the integrand in $k^2$ as 
\begin{eqnarray}
& & \int_0^t d\nu\left[ \exp\left( -2\pi\alpha'k^2\frac{\nu(t-\nu)}t \right)\frac{2\pi}t-2\alpha'k^2G_B^0(\nu,t)\frac{2\pi}t-\partial_\nu^2G_B^0(\nu,t) \right. \nonumber \\ [2mm] 
& & \left. \hspace*{1cm}+2\alpha'k^2\left( \pi\frac{\nu(t-\nu)}t+G_B^0(\nu,t) \right)\partial_\nu^2G_B^0(\nu,t) \right]+{\cal O}(k^4). 
\end{eqnarray}
The integral of the first term of the integrand can be understood, without performing the $\nu$-integration, as the gauge theory amplitude (\ref{app_QFT_gauge1}), as explained in section \ref{sec:PMT}. 
We will evaluate the remaining integrals explicitly. 

First, we consider 
\begin{equation}
\int_0^t d\nu\Bigl( -\partial_\nu^2G_B^0(\nu,t) \Bigr). 
\end{equation}
Formally, we obtain 
\begin{equation}
\int_0^t d\nu\Bigl( -\partial_\nu^2G_B^0(\nu,t) \Bigr)\ =\ -4\pi\sum_{n=1}^\infty e^{-2\pi n\nu}\Big|_{{\nu = 0}}^{{\nu = t}}, 
\end{equation}
where we used the symmetry under the exchange of $\nu$ and $t-\nu$. 
Employing the zeta function regularization, we obtain 
\begin{equation}
-4\pi\sum_{n=1}^\infty e^{-2\pi n\nu}\Big|_{{\nu = 0}}^{{\nu = t}}\ =\ -2\pi+{\cal O}(q). 
\end{equation}

Next, we consider 
\begin{equation}
-\frac{4\pi}t\alpha'k^2\int_0^td\nu\,G_B^0(\nu,t). 
\end{equation}
This is a convergent integral. 
We obtain 
\begin{eqnarray}
-\frac{4\pi}t\alpha'k^2\int_0^td\nu\,G_B^0(\nu,t) 
&=& -\frac{8\pi}t\alpha'k^2\int_0^td\nu\,\log(1-e^{-2\pi\nu}) \nonumber \\ [2mm] 
&=& -\frac4t\alpha'k^2\sum_{n=1}^\infty \frac1{n^2}e^{-2\pi n\nu}\Big|_{{\nu = 0}}^{{\nu = t}} \nonumber \\ [2mm] 
&=& \frac{2\pi^2}{3t}\alpha'k^2+{\cal O}(t^{-1}q). 
\end{eqnarray}

The evaluation of the remaining integrals is a bit subtle. 
We evaluate as follows. 
\begin{eqnarray}
\frac{2\pi}t\alpha'k^2\int_0^td\nu\,\nu(t-\nu)\partial_\nu^2G_B^0(\nu,t) 
&=& -\frac{2(2\pi)^3}t\alpha'k^2\sum_{n=1}^\infty n\int_0^td\nu\,\nu(t-\nu)e^{-2\pi n\nu} \nonumber \\
&=& \alpha'k^2\sum_{n=1}^\infty\left( \frac4{tn^2}-\frac{4\pi}n \right)+{\cal O}(tq) \nonumber \\
&=& \frac{2\pi^2}{3t}\alpha'k^2-4\pi\zeta(1)\alpha'k^2+{\cal O}(tq). \\
2\alpha'k^2\int_0^td\nu\,G_B^0(\nu,t)\partial_\nu^2G_B^0(\nu,t) 
&=& 4(2\pi)^2\alpha'k^2\sum_{m=1}^\infty\sum_{n=1}^\infty \frac nm\int_0^td\nu\,\Bigl( e^{-2\pi m\nu}+e^{-2\pi m(t-\nu)} \Bigr)e^{-2\pi n\nu} \nonumber \\
&=& 8\pi\alpha'k^2\sum_{m=1}^\infty\sum_{n=1}^\infty\frac n{m(n+m)}+{\cal O}(q). 
\end{eqnarray}
We evaluate the last sum as 
\begin{eqnarray}
\sum_{m=1}^\infty\sum_{n=1}^\infty\frac n{m(n+m)} 
&=& \sum_{n=1}^\infty\sum_{m=1}^\infty\left( \frac1m-\frac1{m+n} \right) \nonumber \\
&=& \zeta(0)\zeta(1)-\sum_{N=2}^\infty\sum_{n=1}^{N-1}\frac1N \nonumber \\
&=& -\frac12\zeta(1)-(\zeta(0)-\zeta(1)) \nonumber \\ [2mm] 
&=& \frac12\zeta(1)+\frac12. 
\end{eqnarray}
Using this, we finally obtain 
\begin{equation}
2\alpha'k^2\int_0^td\nu\left( \pi\frac{\nu(t-\nu)}t+G_B^0(\nu,t) \right)\partial_\nu^2G_B^0(\nu,t)\ =\ \left( \frac{2\pi^2}{3t}+4\pi \right)\alpha'k^2+{\cal O}(tq). 
\end{equation}
Note that the divergencies appearing as $\zeta(1)$ in the middle of the calculations cancel among them in the final expression. 

\vspace{1cm}

\section{Bulk propagators and stringy corrections} \label{app_bulk}

\vspace{5mm}

In this paper, we use functions $g_D(\rho)$ to describe stringy corrections. 
These are defined as 
\begin{equation}
g_D(\rho)\ :=\ 2\pi\int_0^\infty ds\,(8\pi^2s)^{-D/2}\exp\left( -\frac{\rho^2}{8\pi s}-2\pi s \right). 
   \label{app_g_D_def}
\end{equation}
This is related to the Green function $G_D(x)$ which satisfies 
\begin{equation}
(-\Delta_D+m^2)G_D(x)\ =\ \delta^D(x), 
   \label{app_partial_diff}
\end{equation}
where $\Delta_D$ is the Laplacian in $\mathbb{R}^D$. 
This can be shown as follows. 
\begin{eqnarray}
G_D(x) 
&=& \int\frac{d^Dp}{(2\pi)^D}\frac{e^{ipx}}{p^2+m^2} \nonumber \\ [2mm] 
&=& 2\pi\int_0^\infty ds\int\frac{d^Dp}{(2\pi)^D}e^{ipx-2\pi (p^2+m^2)s} \nonumber \\ [2mm] 
&=& 2\pi\int_0^\infty ds\,(8\pi^2s)^{-D/2}\exp\left( -\frac{r^2}{8\pi s}-2\pi m^2 s \right) \nonumber \\ [2mm] 
&=& m^{D-2}g_D(mr), 
\end{eqnarray}
where $r:=\sqrt{x^2}$. 

The explicit expression for $g_D(\rho)$ can be obtained from this relation to $G_D(x)$. 
Since $G_D(x)$ is in fact a function of $r$, the equation (\ref{app_partial_diff}) for $r>0$ is reduced to 
\begin{equation}
\left[ \frac{d^2}{dr^2}+\frac{D-1}r\frac d{dr}-m^2 \right]G_D\ =\ 0. 
\end{equation}
Then, the function $\phi_D(r):=r^{D/2-1}G_D(r)$ satisfies 
\begin{equation}
\frac{d^2\phi_D}{d\rho^2}+\frac1{\rho}\frac{d\phi_D}{d\rho}-\left[ 1+\frac{(\frac D2-1)^2}{\rho^2} \right]\phi_D\ =\ 0, 
\end{equation}
where $\rho:=mr$. 
Since $\phi_D(r)$ should decrease for large $r$, we find that $\phi_D(r)$ is proportional to the modified Bessel function $K_{D/2-1}(\rho)$. 
The proportionality constant is fixed by comparing the saddle-point analysis of (\ref{app_g_D_def}) with the asymptotic expansion of $K_{D/2-1}(\rho)$ for large $\rho$. 
As a result, we obtain 
\begin{equation}
g_D(\rho)\ =\ (2\pi)^{-D/2}\rho^{1-D/2}K_{D/2-1}(\rho). 
\end{equation}

\vspace{5mm}

In the calculations of the stringy corrections, $g_D(\rho)$ is used in the following integral formula  
\begin{equation}
\int_0^\infty ds\,s^{-D/2}e^{-2\pi r^2/s}e^{-2\pi ns}\ =\ \frac1{2\pi n}(8\pi^2n)^{D/2}g_D(4\pi\sqrt{n}r). 
\end{equation}
In addition, we use the following formulas 
\begin{eqnarray}
\int_0^\infty ds\,s^{-D/2}e^{-2\pi r^2/s} 
&=& \Gamma({\textstyle \frac D2-1})(2\pi r^2)^{1-D/2}, \\ [2mm] 
\int_0^\infty ds\,s^{-D/2}e^{-2\pi r^2/s}\log s 
&=& \frac d{dx}\int_0^\infty\,s^{-D/2+x}e^{-2\pi r^2/s}\Big|_{x=0} \nonumber \\ [2mm] 
&=& \Gamma({\textstyle \frac D2-1})\Bigl( \log(2\pi r^2)-\psi({\textstyle \frac D2-1}) \Bigr)(2\pi r^2)^{1-D/2}, 
\end{eqnarray}
where $\psi(x)$ is the digamma function. 

By using these formulas for the mass shift (\ref{mass_shift_full}), we obtain 
\begin{eqnarray}
\Delta m^2(r) 
&=& -\frac\xi{\alpha^\prime}\int_{\lambda^2}^\infty ds\,s^{(p-11)/2}e^{-2\pi r^2/s}\frac{\vartheta_{10}(0,is)^4}{\eta(is)^{12}}\Bigl( 2(2\pi)^2r^2s^{-1}-2\pi \Bigr) \nonumber \\ [2mm] 
&=& -16\frac\xi{\alpha^\prime}\int_{\lambda^2}^\infty ds\,s^{(p-11)/2}e^{-2\pi r^2/s}\Bigl( 2(2\pi)^2r^2s^{-1}-2\pi \Bigr) \nonumber \\ [2mm] 
& & -\frac\xi{\alpha^\prime}\sum_{n=1}^\infty c_n\int_{\lambda^2}^\infty ds\,s^{(p-11)/2}e^{-2\pi r^2/s}e^{-2\pi ns}\Bigl( 2(2\pi)^2r^2s^{-1}-2\pi \Bigr), 
\end{eqnarray}
where the coefficients $c_n$ are defined as 
\begin{equation}
\frac{\vartheta_{10}(0,is)^4}{\eta(is)^{12}}\ =\ 16+\sum_{n=1}^\infty c_n\tilde{q}^n. 
\end{equation}
The first integral gives $\Delta m^2_{(0)}(r)$. 
The other integrals can be written in terms of $g_D(\rho)$ as 
\begin{eqnarray}
& & \int_{\lambda^2}^\infty ds\,s^{(p-11)/2}e^{-2\pi r^2/s}e^{-2\pi ns}\Bigl( 2(2\pi)^2r^2s^{-1}-2\pi \Bigr) \nonumber \\ [2mm] 
&=& \frac1n(8\pi^2n)^{(11-p)/2}\Bigl( 2\pi(4\pi\sqrt{n}r)^2g_{13-p}(4\pi\sqrt{n}r)- g_{11-p}(4\pi\sqrt{n}r) \Bigr). 
\end{eqnarray}

\vspace{1cm}

\section{Boundary conditions in Green-Schwarz formalism} \label{boundary_condition_GS} \label{app_GS}

\vspace{5mm}

In this appendix, we derive the boundary condition for an open string attached to a D$p$-brane. 
We discuss in Type IIB string theory. 
The necessary modification for Type IIA case is evident. 
In this appendix, we assume that $x^I$-directions for $I=1,2,\cdots,9-p$ are transverse to the D$p$-brane. 
The supercharges conserved by the D$p$-brane are \cite{Polchinski:1998rr} 
\begin{equation}
Q_\alpha+(\beta^\perp\tilde{Q})_\alpha,
\end{equation}
where 
\begin{equation}
\beta^\perp\ :=\ \prod_{I=1}^{9-p}\Gamma\Gamma^I, \hspace{1cm} \Gamma:=\Gamma^0\cdots\Gamma^9. 
\end{equation}

We rewrite these supercharges in terms of spinors and the gamma matrices $\gamma^i$ $(i=1,\cdots,8)$ for ${\rm Spin}(8)$. 
We use the following representation of the ten-dimensional gamma matrices $\Gamma^\mu$: 
\begin{equation}
\Gamma^0\ =\ \varepsilon\otimes\gamma, \hspace{1cm} \Gamma^i\ =\ I_2\otimes \gamma^i, \hspace{1cm} \Gamma^9\ =\ \sigma_1\otimes \gamma, 
\end{equation}
where 
\begin{equation}
\varepsilon\ :=\ i\sigma_2, \hspace{1cm} \gamma\ :=\ \gamma^1\cdots\gamma^8, 
\end{equation}
and $\sigma_1, \sigma_2$ are the Pauli matrices. 
The explicit form of $\gamma^i$ can be found in \cite{Green:1987sp}. 

In this representation, the matrix $\beta^\perp$ can be written as 
\begin{equation}
\beta^\perp\ =\ I_2\otimes\gamma^{(p)}, \hspace{1cm} \gamma^{(p)}\ :=\ (-1)^{(9-p)/2}\gamma^1\cdots\gamma^{9-p}. 
\end{equation}
Note that $p$ is odd in Type IIB string theory. 
We choose $Q_\alpha$ and $\tilde{Q}_\alpha$ such that they satisfy 
\begin{equation}
\Gamma Q\ =\ Q, \hspace{1cm} \Gamma\tilde{Q}\ =\ \tilde{Q}. 
\end{equation}
Since we have $\Gamma\ =\ \sigma_3\otimes \gamma$, the conserved supercharges can be written as 
\begin{equation}
Q_\alpha+(\beta^\perp\tilde{Q})_\alpha\ =\ \left[
\begin{array}{c}
Q_a+\gamma^{(p)}_{ab}\tilde{Q}_b \\
0 \\
0 \\
Q_{\dot{a}}+\gamma^{(p)}_{\dot{a}\dot{b}}\tilde{Q}_{\dot{b}} 
\end{array}
\right]. 
\end{equation}

\vspace{5mm}

The worldsheet action (\ref{GS_action}) is invariant under the dynamical supersymmetry 
\begin{equation}
\delta X^i\ =\ \frac2{\sqrt{2p^+}}\gamma^i_{a\dot{a}}\bar{\epsilon}^{\dot{a}}S^a, \hspace{1cm} \delta S_a\ =\ -\frac i{2\alpha'\sqrt{2p^+}}\rho^\alpha\epsilon^{\dot{a}}\partial_\alpha X^i\gamma^i_{a\dot{a}}, 
\end{equation}
as well as the kinematical supersymmetry 
\begin{equation}
\delta X^i\ =\ 0, \hspace{1cm} \delta S^a\ =\ \sqrt{2p^+}\eta^a. 
\end{equation}
The corresponding supercurrents are 
\begin{equation}
\left[
\begin{array}{c}
J^\alpha_{\dot{a}} \\ [2mm] 
\tilde{J}^\alpha_{\dot{a}}
\end{array}
\right]\ \propto\ \rho^\beta\rho^\alpha
\left[
\begin{array}{c}
S^a \\ [2mm] 
\tilde{S}^a
\end{array}
\right]\gamma^i_{a\dot{a}}\partial_\beta X^i, \hspace{1cm} 
\left[
\begin{array}{c}
J^\alpha_a \\ [2mm] 
\tilde{J}^\alpha_a
\end{array}
\right]\ \propto\ \rho^0\rho^\alpha 
\left[
\begin{array}{c}
S_a \\ [2mm] 
\tilde{S}^a
\end{array}
\right], 
\end{equation}
where $\rho^0:=\sigma_2$ and $\rho^1:=i\sigma_1$. 
The overall constants are not important in what follows. 
The supercharges are obtained as 
\begin{equation}
Q_{\dot{a}}\ =\ \int_0^\pi d\sigma\,J_{0\dot{a}}, \hspace{1cm} Q_a\ =\ \int_0^\pi d\sigma\,J_{0a}, 
\end{equation}
and similar for $\tilde{Q}_{\dot{a}}$ and $\tilde{Q}_a$. 

We require that the supercharges $Q_{\dot{a}}+\gamma^{(p)}_{\dot{a}\dot{b}}\tilde{Q}_{\dot{b}}$ are conserved in the worldsheet theory. 
The conservation law implies 
\begin{equation}
\frac d{d\tau}\Bigl( Q_{\dot{a}}+\gamma^{(p)}_{\dot{a}\dot{b}}\tilde{Q}_{\dot{b}} \Bigr)\ =\ \Bigl( J_{1\dot{a}}+\gamma^{(p)}_{\dot{a}\dot{b}}\tilde{J}_{1\dot{b}} \Bigr) \Big|_{\sigma=\pi}-\Bigl( J_{1\dot{a}}+\gamma^{(p)}_{\dot{a}\dot{b}}\tilde{J}_{1\dot{b}} \Bigr) \Big|_{\sigma=0}. 
\end{equation}
The linear combination $J_{1\dot{a}}+\gamma^{(p)}_{\dot{a}\dot{b}}\tilde{J}_{1\dot{b}}$ at, say $\sigma=0$, can be written as 
\begin{eqnarray}
J_{1\dot{a}}+\gamma^{(p)}_{\dot{a}\dot{b}}\tilde{J}_{1\dot{b}} 
&\propto& \gamma^l_{\dot{a}a}\partial_\tau X^l\Bigl( S_a-\gamma^{(p)}_{ab}\tilde{S}_b \Bigr)-\gamma^I_{\dot{a}a}\partial_\sigma X^I\Bigl( S_a-\gamma^{(p)}_{ab}\tilde{S}_b \Bigr), 
\end{eqnarray}
where $l=9-p+1,\cdots,8$ labels the directions parallel to the D$p$-brane other than the light cone directions. 
We have used the Neumann/Dirichlet boundary conditions for $X^i$. 
This implies that the boundary condition for the open string attached to the D$p$-brane at $\sigma=0$ is 
\begin{equation}
S_a\ =\ \gamma^{(p)}_{ab}\tilde{S}_b
\end{equation}
at the endpoint. 
The same condition ensures that the other half $Q_a+\gamma^{(p)}_{ab}\tilde{Q}_b$ are also conserved. 

\vspace{5mm}

In section \ref{sec:typeII}, we consider an open string stretched between a D$p$-brane and a $\overline{{\rm D}p}$-brane which are parallel to each other. 
To be more specific, we assume that the open string is attached to the D$p$-brane at $\sigma=0$, and to $\overline{{\rm D}p}$-brane at $\sigma=\pi$. 
In this case, the boundary condition for this open string is 
\begin{equation}
S_a(\tau,0)\ =\ \gamma^{(p)}_{ab}\tilde{S}_b(\tau,0), \hspace{1cm} S_a(\tau,\pi)\ =\ -\gamma^{(p)}_{ab}\tilde{S}_b(\tau,\pi). 
\end{equation}
In the quantization, we use the doubling trick with this boundary condition taken into account. 
As usual, the equations of motion imply 
\begin{equation}
S_a(\tau,\sigma)\ =\ S_a(\tau-\sigma), \hspace{1cm} \tilde{S}_a(\tau,\sigma)\ =\ \tilde{S}_a(\tau+\sigma). 
\end{equation}
We can extend $S_a(\tau,\sigma)$ to the range $-\pi\le\sigma\le0$ by 
\begin{equation}
S_a(\tau,\sigma)\ =\ \gamma^{(p)}_{ab}\tilde{S}_b(\tau,-\sigma). 
\end{equation}
This respects the boundary condition at $\sigma=0$. 
At $\sigma=\pi$, we find 
\begin{equation}
S_a(\tau,\pi)\ =\ -\gamma^{(p)}_{ab}\tilde{S}_b(\tau,\pi)\ =\ -S_a(\tau,-\pi). 
\end{equation}
This implies that the fermionic fields $S^a$ for the open string under consideration obey the anti-periodic or NS boundary condition. 

\vspace{1cm}

\section{Wave function renormalization from DBI action} \label{app_DBI}

\vspace{5mm}

In this appendix, we show a calculation based on Type II supergravity and DBI action for a D$p$-brane which corresponds to the 1-loop wave function renormalization for massless fields on the D$p$-brane. 

First, recall the calculation of the effective potential for a pair of D$p$-branes \cite{Polchinski:1998rr}. 
See also \cite{Bachas:1998rg}. 
We calculate the tree amplitudes for the exchange of massless bulk fields between the D$p$-branes. 
The coupling vertices of the D$p$-brane and the bulk fields are determined from DBI action 
\begin{equation}
S_p\ =\ -T_p\int d^{p+1}\xi\,e^{-\Phi}\sqrt{-\det\Bigl( g+B+2\pi\alpha'F \Bigr)}. 
   \label{app_DBI_action}
\end{equation}
From this we obtain 
\begin{equation}
\mbox{graviton coupling}\ :\ -\frac12T_p\eta^{\mu\nu}, \hspace{1cm} \mbox{dilaton coupling}\ :\ -\frac{p-3}4T_p, 
   \label{app_DBI_non-derivative}
\end{equation}
by expanding $S_p$ in terms of the fluctuations of $g_{\mu\nu}$ and $\Phi$ around the flat background with the vanishing vev for $\Phi$. 
The B-field does not couple directly to the D$p$-brane, and therefore it is not relevant for the effective potential. 
The couplings to R-R fields are crucial. 
However, for the calculations corresponding to the wave function renormalization, those couplings do not contribute, as we will explain below. 

The couplings (\ref{app_DBI_non-derivative}) do not depend on the derivative of the fields on the D$p$-brane. 
As a result, the exchange amplitudes give an effective potential which does not include derivatives. 
On the other hand, the wave function renormalization comes from ${\cal O}(k^2)$ terms in the 1-loop amplitudes. 
The corresponding exchange amplitude is depicted in figure \ref{fig_closed_channel}. 
To obtain such exchange amplitudes, we need to employ coupling vertices including two worldvolume fields, two derivatives, and one bulk field, derived from DBI action (\ref{app_DBI_action}). 

\vspace{5mm}

\subsection{Gauge fields}

\vspace{5mm}

As a warmup, let us consider the wave function renormalization for gauge fields. 
As in subsection \ref{sec:typeIIppbar}, we consider a D$p$-$\overline{{\rm D}p}$ pair and the gauge fields are propagating on the D$p$-brane. 
We find that the coupling vertices for the D$p$-brane are 
\begin{eqnarray}
\mbox{graviton coupling} &:& -(\pi\alpha^\prime)^2T_p\left( -2F^\mu{}_\alpha F^{\nu\alpha}+\frac12\eta^{\mu\nu}F^2 \right), \\ [2mm] 
   \label{app_DBI_derivative-graviton}
\mbox{dilaton coupling} &:& -\frac{p-7}4(\pi\alpha^\prime)^2T_pF^2. 
   \label{app_DBI_derivative_dilaton}
\end{eqnarray}
For the $\overline{{\rm D}p}$-brane we choose (\ref{app_DBI_non-derivative}). 

The couplings of R-R fields to the D$p$-brane are given by the Chern-Simons-like terms \cite{Polchinski:1998rr} 
\begin{equation}
S_{\rm CS}\ =\ i\mu_p\int \exp\Bigl( B+2\pi\alpha'F \Bigr)\wedge\sum_qC_q 
\end{equation}
added to (\ref{app_DBI_action}). 
More precisely, there are also couplings to the curvature \cite{Green:1996dd,Cheung:1997az}. 
The D$p$-brane couples to $(p+1)$-form field $C_{p+1}$ via a non-derivative coupling. 
We also need coupling vertices including two worldvolume gauge fields and two derivatives in our calculation of the wave function renormalization. 
However, such terms only allow the couplings of the D$p$-brane to $(p-3)$-form field $C_{p-3}$. 
Then, neither $C_{p+1}$ nor $C_{p-3}$ propagates in the amplitude for figure \ref{fig_closed_channel}. 
Therefore, the R-R fields do not contribute to the wave function renormalization. 

Note that the irrelevance of the R-R fields in our calculation actually implies what the resulting amplitude should be.
Namely, no propagation of R-R fields means that the R-R charges of the D-branes are irrelevant. 
Therefore, the amplitude for D$p$-$\overline{{\rm D}p}$ pair should be the same as the amplitude for a pair of D$p$-branes. 
Since the string 1-loop amplitude on BPS D$p$-branes vanishes, the exchange amplitudes for D$p$-$\overline{{\rm D}p}$ pair at the level of supergravity should also vanish. 

This can be checked explicitly by using the coupling vertices obtained above and the propagators 
\begin{eqnarray}
\mbox{graviton} &:& \left( \eta_{MK}\eta_{NL}+\eta_{ML}\eta_{NK}-\frac14\eta_{MN}\eta_{KL} \right)\Delta(x), 
   \label{app_graviton_propagator} \\ [2mm] 
\mbox{dilaton} &:& \Delta(x), 
\end{eqnarray}
where $M,N$ etc. run from 0 to 9, and $\Delta(x)$ is the propagator of a scalar field in ten dimensions. 
The graviton exchange is given by 
\begin{eqnarray}
& & -(\pi\alpha^\prime)^2T_p\left( -2F^\mu{}_\alpha F^{\nu\alpha}+\frac12\eta^{\mu\nu}F^2 \right)\left( \eta_{\mu\rho}\eta_{\nu\sigma}+\eta_{\mu\sigma}\eta_{\nu\rho}-\frac14\eta_{\mu\nu}\eta_{\rho\sigma} \right){\Delta(x)}\left( -\frac12T_p\eta^{\rho\sigma} \right) \nonumber \\ [2mm] 
&=& -\frac{(p-3)(p-7)}{16}(\pi\alpha')^2T_p^2{\Delta(x)}F^2. 
\end{eqnarray}
The dilaton exchange is given by 
\begin{equation}
\frac{(p-3)(p-7)}{16}(\pi\alpha^\prime)^2{T_p^2\Delta(x)}F^2. 
\end{equation}
We find that the amplitudes cancel completely, as expected. 

\vspace{5mm}

\subsection{Transverse scalars}

\vspace{5mm}

Next, we consider the wave function renormalization for transverse scalars $\phi^k$, where $k$ runs from $p+1$ to 9. 
For this purpose, we need coupling vertices for $\phi^k$ and the bulk fields, but it is not apparent in the action (\ref{app_DBI_action}). 
It becomes explicit if we recall that the metric $g$ in (\ref{app_DBI_action}) is in fact the induced metric on the D$p$-brane. 
This means that we should replace $g$ with 
\begin{equation}
g_{\mu\nu}+\partial_\mu\phi^k\partial_\nu\phi^lh_{kl}, 
\end{equation}
where $h_{kl}$ are the components of the target space metric for transverse directions. 
Therefore, the coupling vertices relevant for our calculation are obtained from 
\begin{equation}
-\frac12T_p\int d^{p+1}\xi\,e^{-\Phi}\sqrt{-g}g^{\mu\nu}\partial_\mu\phi^k\partial_\nu\phi^lh_{kl} 
\end{equation}
which is obtained by expanding (\ref{app_DBI_action}). 
Note that $h_{kl}$ are also propagating although they do not couple to the $\overline{{\rm D}p}$-brane, since the graviton propagator (\ref{app_graviton_propagator}) allows the mixing of $h_{kl}$ and $g_{\mu\nu}$. 
We obtain the following coupling vertices 
\begin{eqnarray}
\mbox{graviton}\ g_{\mu\nu} &:& \frac12T_p\left( \partial^\mu\phi^k\partial^\nu\phi_k-\frac12\eta^{\mu\nu}(\partial\phi)^2 \right), \\ [2mm] 
\mbox{graviton}\ h_{kl} &:& -\frac12T_p\,\partial_\mu\phi^k\partial^\mu\phi^l, \\ [2mm] 
\mbox{dilaton} &:& -\frac{p-3}8T_p(\partial\phi)^2. 
\end{eqnarray}

Now we can calculate the exchange amplitudes. 
The exchange of $g_{\mu\nu}$ is given by 
\begin{eqnarray}
& & \frac12T_p\left( \partial^\mu\phi^k\partial^\nu\phi_k-\frac12\eta^{\mu\nu}(\partial\phi)^2 \right)\left( \eta_{\mu\rho}\eta_{\nu\sigma}+\eta_{\mu\sigma}\eta_{\nu\rho}-\frac14\eta_{\mu\nu}\eta_{\rho\sigma} \right){\Delta(x)}\left( -\frac12T_p\eta^{\rho\sigma} \right) \nonumber \\ [2mm] 
&=& -\frac{(p-7)(p-1)}{32}T_p^2{\Delta(x)}(\partial \phi)^2. 
\end{eqnarray}
The mixing of $g_{\mu\nu}$ and $h_{kl}$ gives 
\begin{equation}
-\frac12T_p\,\partial_\mu\phi^k\partial^\mu\phi^l\left( -\frac14\delta_{kl}\eta_{\rho\sigma} \right){\Delta(x)}\left( -\frac12T_p\eta^{\rho\sigma} \right)\ =\ -\frac{p+1}{16}T_p^2{\Delta(x)}(\partial\phi)^2. 
\end{equation}
The dilaton exchange is given by 
\begin{equation}
\frac{(p-3)^2}{32}T_p^2{\Delta(x)}(\partial\phi)^2. 
\end{equation}
These three contributions cancel among them completely, as expected. 

\vspace{1cm}

\section{A modified setup in Type II string theory} \label{modified_SS}

\vspace{5mm}
 
\begin{figure}
\begin{center}
\includegraphics{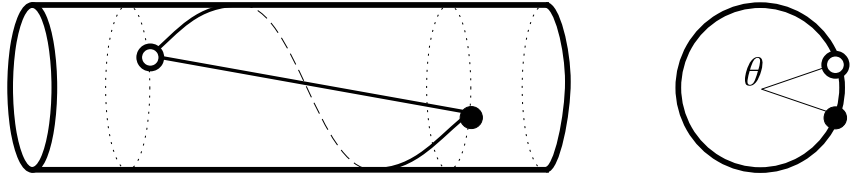}
\end{center}
\caption{Schematic picture of the target space-time considered in appendix \ref{modified_SS}. 
The open circle represents a D$p$-brane, and the closed circle represents a $\overline{{\rm D}p}$-brane, both localized in the cylinder directions. 
The solid lines connecting two circles represent open strings stretched between two D-branes with the winding numbers 0 and 1. 
The relative angle between the D-branes is denoted by $\theta$. }
   \label{fig_SS}
\end{figure}

In this appendix, we consider modifications of the D-brane setup discussed in subsection \ref{sec:typeIIppbar}. 
We put the D$p$-$\overline{{\rm D}p}$ pair on a cylinder, as in figure \ref{fig_SS}. 
For this setup, we need to take into account the windings of the stretched string. 
Let $w$ be the winding number of the string. 
The zero mode contributions to $L_0$ from the string tension becomes 
\begin{equation}
\frac1{4\pi^2\alpha^\prime}\Bigl( l^2+(R\theta+2\pi Rw)^2 \Bigr)\ =\ r^2+\frac{R^2}{\alpha^\prime}w^2+\frac{R^2\theta}{\pi\alpha^\prime}w+\frac{R^2\theta^2}{4\pi^2\alpha^\prime}, 
\end{equation}
where $R$ is the radius of the $S^1$ direction. 
Then, the mass shift in this setup is obtained from (\ref{mass_shift_full}) by inserting 
\begin{equation}
\sum_{w\in\mathbb{Z}}\exp\left[ -2\pi t\left( \frac{R^2}{\alpha^\prime}w^2+\frac{R^2\theta}{\pi\alpha^\prime}w+\frac{R^2\theta^2}{4\pi^2\alpha^\prime} \right) \right]\ =\ \exp\left( -\frac{R^2\theta^2}{2\pi\alpha^\prime}t \right)\vartheta_{00}\left( i\frac{R^2\theta}{\pi\alpha^\prime}t,i\frac{2R^2}{\alpha^\prime}t \right) 
   \label{brane_with_angle}
\end{equation}
in the integral. 

Recall that a hierarchical mass spectrum is realized when there is no contributions from massless closed string exchanges in (\ref{mass_shift_full}). 
In order to examine the masses of the closed string modes, we perform the modular transformation for the 
right-hand side of (\ref{brane_with_angle}). 
As a result, we obtain 
\begin{equation}
\sqrt{\frac{\alpha^\prime s}{2R^2}}\vartheta_{00}\left( \frac\theta{2\pi},i\frac{\alpha^\prime}{2R^2}s \right)\ =\ \sqrt{\frac{\alpha^\prime s}{2R^2}}\sum_{m\in\mathbb{Z}}e^{im\theta}\exp\left( -\frac{\pi\alpha^\prime}{2R^2}m^2 s \right). 
   \label{app_zeromode_KK}
\end{equation}
Since the $m=0$ term gives the massless closed string modes, the mass spectrum in this D-brane setup is not hierarchical. 

\begin{figure}
\begin{center}
\includegraphics{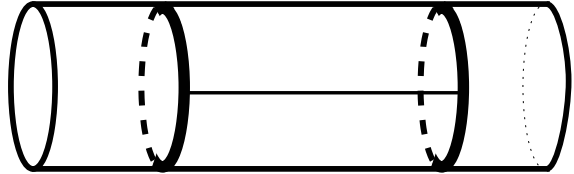}
\end{center}
\caption{The T-dual setup. 
The thick solid line on the left represents the D$(p+1)$-brane, and the right one represents the $\overline{{\rm D}(p+1)}$-brane. 
The solid line in the middle represents a stretched string which can move along the $S^1$ direction. }
   \label{fig_T-dual}
\end{figure}

Note that there are phases $e^{im\theta}$ in the sum. 
The origin of these phases can be understand more easily in the T-dual setup, depicted in figure \ref{fig_T-dual}. 
In this setup, stretched strings can no longer wind around $S^1$. 
Instead, they can have KK momentum along the $S^1$. 
The angle $\theta$ describing the relative position of the D-branes in the $S^1$ direction is encoded in the T-dual setup into the Wilson line on the $\overline{{\rm D}(p+1)}$-brane. 
As a string state with a non-zero KK momentum goes around the $S^1$, the endpoint of the string attached to the $\overline{{\rm D}(p+1)}$-brane receives a holonomy. 
This is the phase appearing the sum (\ref{app_zeromode_KK}). 

It is curious to ask what happens when similar phases are introduced in the sum over winding numbers for figure \ref{fig_SS}. 
If this is justified, then the theta function we should insert into $\Delta m^2(r)$ becomes 
\begin{equation}
\sum_{w\in \mathbb{Z}}e^{i\varphi w}\exp\left( -2\pi\frac{R^2}{\alpha^\prime}w^2t \right)\ =\ \vartheta_{00}\left( \frac\varphi{2\pi},i\frac{2R^2}{\alpha^\prime}t \right), 
\end{equation}
where we have set $\theta=0$ for simplicity. 
The modular transformation of this theta function gives 
\begin{eqnarray}
& & \sqrt{\frac{\alpha^\prime s}{2R^2}}\exp\left( -\frac{\alpha^\prime\varphi^2}{8\pi R^2}s \right)\vartheta_{00}\left( -i\frac{\alpha^\prime\varphi}{4\pi R^2}s,i\frac{\alpha^\prime}{2R^2}s \right) \nonumber \\ [2mm] 
&=& \sqrt{\frac{\alpha^\prime s}{2R^2}}\sum_{m\in\mathbb{Z}}\exp\left[ -\frac{\pi\alpha^\prime}{2R^2}\left( m-\frac\varphi{2\pi} \right)^2s \right]. 
   \label{app_zeromode_winding}
\end{eqnarray}
This shows that the KK momenta of the closed string modes are non-zero for generic $\varphi$. 
In particular, there are no massless closed string states exchanged between the D$p$-branes, resulting in a hierarchical mass spectrum on the D-brane for {\it general} $p$. 
Possibilities of inserting phases in the topological sum was discussed in \cite{Seiberg:2010qd}. 
Indeed, the resulting sum looks quite similar to the one we obtain for the $\theta$-vacuum in Yang-Mills theory in the large $N$ limit \cite{Witten:1998uka}. 

It is known \cite{Polchinski:1998rq} that winding numbers may produce phases when a constant B-field is introduced as a background. 
Recall that the dependence of the worldsheet action on the B-fied is proportional to 
\begin{equation}
i\int d^2\sigma\,\partial_\tau X^\mu\partial_\sigma X^\nu B_{\mu\nu}. 
\end{equation}
This shows that, in order to produce non-trivial phases depending on the winding number, we also need a KK momentum along a perpendicular direction such that the above integral becomes non-vanishing. 
Then, we should consider the setup in figure \ref{fig_T-dual} rather than the one in figure \ref{fig_SS}. 
We find that the theta function for figure \ref{fig_T-dual} with a constant B-field turns out to be of the form (\ref{app_zeromode_KK}) with $\theta$ given by the B-field, not of the form (\ref{app_zeromode_winding}). 
We conclude that the desired phases cannot be introduced in this manner.

\end{document}